\documentclass[12pt]{article}

\usepackage{amsmath}
\usepackage{amssymb}
\usepackage{natbib}
\usepackage[ruled]{algorithm2e}

\usepackage{times}
\usepackage{bm}

\usepackage{paralist} 
\usepackage{mathrsfs}
\usepackage{multirow}
\usepackage{graphicx} 
\usepackage{subfig}
\usepackage{enumitem}
\usepackage{array,booktabs}
\usepackage{float}

\usepackage{authblk}
\usepackage{amsthm}
\newtheorem{theorem}{Theorem}
\newtheorem{lemma}{Lemma}

\newtheorem{definition}{Definition}
\newtheorem{remark}{Remark}
\usepackage[toc,page,title,titletoc,header]{appendix}

\newcommand{\essinf}{{\mathrm{ess}\inf}}

\newcommand{\rd}{\,\mathrm{d}}

\newcommand{\Proj}{{\mathbf P}}
\newcommand{\EP}{\,\mathbb{P}}
\newcommand{\EE}{\,\mathbb{E}}
\newcommand{\bsb}{\boldsymbol}
\newcommand{\tran}{{\mathsf{T}}}
\newcommand{\bsbX}{{\boldsymbol{X}}}
\newcommand{\bsbx}{{\boldsymbol{x}}}
\newcommand{\bsby}{{\boldsymbol{y}}}

\newcommand{\bsbT}{{\boldsymbol{T}}}

\newcommand{\bsbW}{{\boldsymbol{W}}}
\newcommand{\bsbI}{{\boldsymbol{I}}}

\newcommand{\bsbZ}{{\boldsymbol{Z}}}

\newcommand{\bsbY}{{\boldsymbol{Y}}}
\newcommand{\bsbB}{{\boldsymbol{B}}}

\newcommand{\bsbS}{{\boldsymbol{S}}}

\newcommand{\bsbDelta}{{\boldsymbol{\Delta}}}
\newcommand{\bsbv}{{\boldsymbol{v}}}
\newcommand{\bsbSig}{{\boldsymbol{\Sigma}}}
\newcommand{\bsbxi}{{\boldsymbol{\xi}}}
\newcommand{\bsbXi}{\boldsymbol{\Xi}}
\newcommand{\bsbD}{{\boldsymbol{D}}}
\newcommand{\bsbF}{{\boldsymbol{F}}}
\newcommand{\bsbU}{{\boldsymbol{U}}}
\newcommand{\bsbV}{{\boldsymbol{V}}}

\newcommand{\bsba}{{\boldsymbol{\alpha}}}
\newcommand{\bsbA}{{\boldsymbol{A}}}

\newcommand{\bsbC}{{\boldsymbol{C}}}

\newcommand{\bsbE}{{\boldsymbol{E}}}

\newcommand{\bsbdelta}{{\boldsymbol{\delta}}}

\newcommand{\bsbs}{{\boldsymbol{s}}}

\DeclareMathOperator{\vect}{\mbox{vec}\,}

\usepackage{setspace}

\begin{document}

%
%


\title{Selective Factor Extraction in High Dimensions}
\author{Yiyuan She\\Department of Statistics, Florida State University}
\date{}
\maketitle

%

\begin{abstract}
This paper studies  simultaneous  feature selection and   extraction in  supervised and unsupervised learning. We propose and investigate  selective reduced rank regression  for constructing optimal  explanatory factors from a parsimonious  subset of input features. The proposed estimators  enjoy  sharp oracle inequalities, and with  a predictive information criterion  for model selection, they  adapt  to  unknown sparsity by controlling      both rank and row support   of the coefficient matrix.       A class of   algorithms is  developed that can    accommodate various convex and nonconvex   sparsity-inducing penalties, and can be used for    rank-constrained variable screening  in   high-dimensional multivariate data. The paper also  showcases   applications in macroeconomics  and computer vision to demonstrate how low-dimensional data structures can be effectively captured by joint variable selection and projection.
\end{abstract}

%
%
%

\section{Introduction}
\label{sec:intro}
Modern statistical applications may involve many variables.  Principal component analysis  \citep{hotelling}  offers  a popular means of  dimension reduction, and reduced rank regression extends it to supervised learning
\citep{And51} by solving the problem $\min_{\bsbB\in \mathbb R^{p\times m}} \|\bsbY - \bsbX \bsbB\|_F^2$ subject to $r(\bsbB)\leq r$, where $\bsbY\in \mathbb R^{n\times m}$ and $\bsbX\in \mathbb R^{n\times p}$ are   response and predictor matrices,    $r(\bsbB)$ denotes the rank of $\bsbB$, and $\|\cdot \|_F$ is the Frobenius norm. Reduced rank regression   provides a  low-dimensional projection  space to view and analyze multivariate data, and finds widespread applications in  machine learning,   econometrics, and  finance  \citep{ReinVelu,izenbook}. In fact, once an estimate $ \bsbB$ of rank $r$ is obtained, we can   write    $ \bsbB = \bsbB_1 \bsbB_2^{\tran}$ for   $\bsbB_1\in \mathbb R^{p\times r}$, $\bsbB_2\in \mathbb R^{m\times r}$. This suggests that    $r$  factors    can be  constructed  by  $\bsbX\bsbB_1$ from   $p$  predictors  to  explain all  response variables. The number of  factors required in real applications is often much smaller than the number of input $x$-variables.
Unfortunately, the loading matrix $\bsbB_1$ obtained from    reduced rank regression  typically involves {all} predictors.
In high-dimensional data analysis,         factors constructed  from a  small  subset of variables are much more interpretable;   we call this  {selective} factor extraction. Correspondingly, the coefficient matrix  is desired to have  {both}  low rank and row-wise sparsity. To capture    the two types of structural parsimony simultaneously,  joint regularization must be applied, which  adds       nontrivial  difficulties in the theoretical analysis   and numerical computation of the associated estimators,  but     leads to reduced  errors  compared with rank reduction or variable selection alone.

In  the unsupervised setting when $\bsbX =\bsbI$,  selective factor extraction is closely related to   sparse principal component analysis. See, for example,  \cite{spapca}, \cite{shenhuang}, \cite{pmd}, \cite{johnstoneLu2009}  and \cite{Ma2013}.
 Most  of these algorithms seek sparse     loading vectors separately,  and progress  sequentially. The     loading matrix obtained may lack optimality  and  contain too many   variables. To    ensure  dimension reduction even when constructing a number of factors, we will formulate the problem as a whole and pursue joint sparsity across all loading vectors. This   turns out be  particularly helpful in  rank-constrained variable screening. 

 There is less work on simultaneous variable selection and rank reduction in the supervised setting. See  \cite{BSW12},   \cite{ChenHuang12}, \cite{Chen2012jrssb}, \cite{Ma04112014} and a recent report by \cite{masun14}.  Many theoretical and computational  questions   remain open.
Our main contributions are threefold.
First, we are able to  provide a    unified treatment for various  penalties in the reduced rank model,  and successfully build sharper  oracle inequalities than those in the literature   \citep{bickel09,lounici-2010,CandPlan}.   Our results indicate that  for         joint  variable selection and rank reduction, the   error rates and parameter choices previously obtained   are   suboptimal.
Second, we develop  a computational framework with    guaranteed convergence, where    any   thresholding rule can be applied.    The   algorithms   adapt to  reduced rank   variable screening   in  very high dimensions.
Third, we come up with a new information   criterion for parameter tuning. To the best of our knowledge, this is the first    sound        criterion  with minimax optimality for selecting among      sparse and/or   rank-deficient models. 

In the rest of the paper, the following notation and symbols will be  used. Given a matrix $\bsbA=(\bsba_1, \ldots, \bsba_p)^\tran \in \mathbb R^{p\times m}$,   $\|\bsbA\|_F$ and $\|\bsbA\|_2$  denote its Frobenius norm and spectral norm, respectively. We   define the   $(2,1)$-norm  of $\bsbA$ by $\|\bsbA\|_{2,1} = \sum_{j=1}^p  \|\bsba_j\|_2 $, and use    $\|\bsbA\|_{2,0}=\sum_{j=1}^p 1_{\|\bsba_{j}\|\ne \bsb0}$ to characterize  the number of nonzero rows in $\bsbA$.   The standard vectorization   of $\bsbA$  is denoted by  $\vect(\bsbA)$. We use  $\bsbA(\mathcal J, \mathcal I)$ to  denote a submatrix of $\bsbA$ with  rows and columns indexed by $\mathcal J$ and $\mathcal I$, respectively, and occasionally abbreviate $\bsbA(\mathcal J, )$ to $\bsbA_{\mathcal J}$.
The set of column-orthogonal matrices of size $m\times r$ is denoted by $ \mathbb O^{m\times r}=\{\bsbV\in \mathbb R^{m\times r}: \allowbreak\bsbV^{\tran} \bsbV=\bsbI\}$. Finally,   $C$ and $c$ are used to denote   constants which are not necessarily the same at each occurrence.


\section{Simultaneous Rank Reduction and Variable Selection}

\label{sec:pcas}
\subsection{ Selective reduced rank regression}
Picking  only pertinent dimensions  is the key to enhance  interpretability of factors   in high dimensions. In a multi-factor model, power to select  requires eliminating the   nuisance   variables  from the construction of   factors. 
To state a general framework, we assume that a response matrix    $\bsbY \in {\mathbb R}^{n\times m}$ is available, in addition to   a predictor matrix   $\bsbX \in {\mathbb R}^{n\times p}$, both centered column-wise.   Let $\bsbB$ denote the  coefficient matrix  $\bsbB=(\bsb{b}_1, \ldots, \bsb{b}_p)^{\tran}=(b_{j,k}) $. To provide  concurrent rank reduction and feature selection, a possible   optimization criterion is     $\min_{\bsbB\in \mathbb R^{p\times m}}
\| \bsbY - \bsbX \bsbB \|_F^2 + {\lambda_1^2} r(\bsbB)  + {\lambda_2^2} \| \bsbB \|_{2,0} 
$.   But   the penalized form does not seem to enjoy low errors  in either theory or   practice.  We  propose the following form of rank-constrained variable selection
\begin{align}
\min_{\bsbB\in \mathbb R^{p\times m}} \frac{1}{2} \| \bsbY - \bsbX \bsbB \|_F^2 + \sum_{j=1}^p P(\|\bsb{b}_j\|_2; \lambda) \quad \mbox{subject to} \quad r(\bsbB) \leq r, \label{sel-rrr}
\end{align}
where  $P$ is a sparsity-promoting penalty,  possibly nonconvex. We call   \eqref{sel-rrr}     selective reduced-rank regression.  Imposing element-wise sparsity on $\bsbB$, though valid as a   regularization approach, does not seem to have much meaning in applications. We will introduce a different sparse reduced rank regression  in \eqref{sparrr} by sparsifying a component of $\bsbB$.

There is a variety of  choices for the penalty function. The
  popular group  $\ell_1$-norm function,   $\lambda \| \bsbB\|_{2,1}$   \citep{Yuan}, leads to    the   rank-constrained group lasso,     although the      group $\ell_0$ penalty, $(\lambda^2/2)  \| \bsbB\|_{2,0}$,   is arguably better in promoting sparsity \citep{BSW12,ChenHuang12}.
Group  versions of   the nonconvex  penalties   proposed by  \cite{FanLi},  \cite{mplus}, and  \cite{Zhang10} can   also be applied.

Solving the selective reduced rank regression problem helps uncover  factors that reduce   model complexity.   Given a selective reduced rank regression estimate $\hat \bsbB$, we can    use its  column space to make a new model matrix $\bsbZ=\bsbX \bsbU \bsbD$, where   $\bsbU$, $\bsbD$, and  $\bsbV$ are obtained from  the  singular value decomposition $\hat\bsbB=\bsbU \bsbD \bsbV^{\tran}$. The new design  has   $r$ columns and  involves  only  a small subset of the $x$-variables.  This is called     {Type-I  factor extraction}. 
An alternative is to  decompose $\bsbX \hat \bsbB$. Concretely, let   $\hat\bsbB^{\tran}\bsbX^{\tran}\bsbX\hat \bsbB = \bsbV \bsbD \bsbV^{\tran}$ be the spectral decomposition of $\hat\bsbB^{\tran}\bsbX^{\tran}\bsbX\hat \bsbB$. Then
$
\bsbZ = \bsbX \hat\bsbB \bsbV
$ provides $r$ factors,   called       {Type-II extraction} or post-decorrelation,  since   the $z$-variables are uncorrelated with each other.
 QR decomposition can be used  for   efficiency reasons in either case.  The two types of factor extraction   are   not equivalent   in general, but coincide when $\hat\bsbB$ is the solution to     reduced rank regression. Because $r\ll p$, a more sophisticated model can be built on the   factors with relative ease.

\subsection{Oracle inequalities }
\label{subsec:oracleineq}
We show some non-asymptotic oracle inequalities to reveal the theoretical benefits of    selective reduced rank regression.    For clarity, we use the  group $\ell_0$ and group $\ell_1$   penalties to exemplify the  error rate.   For $\bsbB=(\bsb{b}_1,\ldots, \bsb{b}_p)^\tran$, define    $\mathcal J(\bsbB)= \{j: \bsb{b}_j \neq \bsb{0}\}$ and   $J(\bsbB) = | \mathcal J(\bsbB)|=\|\bsbB\|_{2,0}$. 
\begin{theorem}\label{thm:oracleerr}

Let    $\bsbY = \bsbX \bsbB^* +  \bsbE$, with  all entries of $\bsbE$  independent and identically distributed  as $ \mathcal N(0, \sigma^2)$.

(i) Let $\hat\bsbB$ be a selective reduced rank regression estimator that minimizes  $\| \bsbY - \bsbX \bsbB\|_F^2 +  {\lambda^2} \|\bsbB\|_{2,0}$ subject to $r(\bsbB)\leq r$. Then, under $\lambda = A \sigma (r + \log p)^{1/2}$ where $A$  is a large enough constant,  the following oracle inequality holds
for any $\bsbB\in \mathbb R^{p\times m}$ with $r(\bsbB)\leq r$,
\begin{align}
 \EE(\| \bsbX \hat \bsbB - \bsbX \bsbB^*\|_F^2)   \lesssim
   \| \bsbX \bsbB - \bsbX \bsbB^*\|_F^2 +\lambda^2 J(\bsbB) +(m-r)r \sigma^2 + \sigma^2.  \label{selrrroraineq}
\end{align}
Here,      $\lesssim$   means that the inequality  holds up to a multiplicative  constant.

(ii) In the $\ell_1$ case, let   $\hat \bsbB = \arg \min_{\bsbB: r(\bsbB)\leq r} \| \bsbY - \bsbX \bsbB\|_F^2/(2\|\bsbX\|_2) + \lambda \|\bsbB\|_{2,1}$  where  $\lambda$ is  as in (i). Then     $\EE(\| \bsbX \hat \bsbB - \bsbX \bsbB^*\|_F^2)    \lesssim
   \| \bsbX \bsbB - \bsbX \bsbB^*\|_F^2 + K^2 \lambda^2 J(\bsbB) +(m-r)r \sigma^2 + \sigma^2$    holds for  any $\bsbB\in \mathbb R^{p\times m}$  with $r(\bsbB)\leq r$, provided that       $\bsbX$ satisfies
\begin{align}
(1+\vartheta) \|\bsbX\|_2 \| \bsbDelta_{\mathcal J}\|_{2,1} \le K   |\mathcal J|^{1/2} \| \bsbX \bsbDelta\|_F +  \|\bsbX\|_2 \| \bsbDelta_{\mathcal J^c}\|_{2,1}, \ \ \bsbDelta\in \mathbb R^{p\times m} \label{compregcond0} \end{align}
where $\mathcal J= \mathcal J(\bsbB)$, $K\ge 0$, and $\vartheta$ is a   positive constant.
\end{theorem}



The proof    given in the Appendices   can deal with various     penalties in a universal way. For example,  the oracle inequality \eqref{selrrroraineq}  applies to any   $P(\cdot;\lambda)$  that takes $\lambda$ as the threshold and  satisfies $P_H(\theta; \lambda) \le P(\theta;\lambda)\le C\lambda^2$, where  $P_H(\theta; \lambda)= (-\theta^2/2+\lambda |\theta|)1_{|\theta|<\lambda} +(\lambda^2/2) 1_{|\theta|\geq \lambda}
$. Examples   includes     the smoothly clipped absolute deviation penalty \citep{FanLi},   the minimax concave penalty   \citep{mplus}  and   the capped $\ell_1$ penalty  \citep{Zhang10}. Similarly, the  result in part (ii) of Theorem  \ref{thm:oracleerr} holds for  any   sub-additive penalty  that is sandwiched between $P_H(\theta;\lambda)$ and   $P_1(\theta;\lambda)=\lambda |\theta|$.  The  $\ell_{ p}$ penalties 
 $P(\theta; \lambda)=  {(2-2 p)}^{1- p}(2- p)^{ p-2}  \lambda^{2- p}|\theta|^{ p}$   where  $0< p<1$ are   particular instances.
Moreover, condition \eqref{compregcond0} is less demanding than some common regularity assumptions  \citep{van2009conditions,Shetheta}, and  we   do not   require  $\|\bsbX\|_2$ to be bounded above  by   $C {n}^{1/2}$. 

  Let $r^* = r(\bsbB^*)$,   $J^*=|\mathcal J(\bsbB^*) |$. According to  \eqref{selrrroraineq}, simply taking  $r=r^*$, $\bsbB = \bsbB^*$  so that  the bias term $\| \bsbX \bsbB - \bsbX \bsbB^*\|_F^2$  disappears, we get  a prediction error bound of  order
\begin{align}(J^*+m-r^*)r^*+J^*\log p, \label{rateofselrrr}
\end{align}  omitting     $\sigma^2$ and constant factors.
The  bias term    makes  the error bound applicable   to coefficient matrices that are     {approximately} {row-sparse}.
In Section \ref{subsec:screening}, we will  see that when it is  difficult to provide a proper rank value,   the    predictive information criterion  can be used to tune  $r$ to guarantee     the same low   error rate.


A comparison between Theorem \ref{thm:oracleerr} and  some existing non-asymptotic results  follows.  \cite{Weihuang} and \cite{lounici-2010} showed that for group lasso, the prediction error    is of  the order $J^*m + J^*\log p$.
Since          $(J^*+m-r^*)r^*+J^*\log p\lesssim J^*m + J^*\log p$, selective  rank reduction is uniformly better,  and  the    performance gain is   dramatic for low-rank models.
\cite{BSW12}  obtained  an error rate for the rank-constrained group lasso at   $J^* r^* \log p+mr^*  $. Their rate is, however,     suboptimal:  when $r^*$ and  $J^*$ are comparable, their  error bound  is  of  the order   $ J^{* 2}  \log p+J^* m $, while   \eqref{rateofselrrr} gives    ${J^*}  \log p + J^* m  $. 
 Bunea et al.  also required a multivariate restricted eigenvalue assumption that is more restrictive than \eqref{compregcond0}.
Compared with     low-rank matrix estimation \citep{Recht2010,BuneaSheWeg}, which has an error rate of  $mr+q r$ with $q=r(\bsbX)$, our result does  not always show an improvement,  
    because   only large values of $A$ are considered  in Theorem \ref{thm:oracleerr} to secure selectivity.   Practically  there will be  {no} performance loss,  because  selective reduced rank regression degenerates to    reduced rank regression when $\lambda=0$. 


\section{Parameter Tuning and Model Comparison 
}
\label{sec:pic}
\subsection{A predictive information criterion}
Selective reduced rank regression has  two regularization parameters, $\lambda$ and $r$, to control the row support and rank of the model. Conventional   tuning methods are not  satisfactory  in  our  experience, and indeed they all  lack  theoretical support in the    sparse and rank-deficient setting.  We will propose a novel information  criterion  from the perspective of  {predictive learning} \citep{ESL2}, namely,   the best model  should give the smallest  prediction error among all candidate models. Unlike   consistent variable selection or rank selection, such a principle does not require high   signal-to-noise ratios to work. 





  To make our results more general, the noise matrix is assumed to have  sub-Gaussian marginal tails in this section.
   A  random variable  $\xi$ is sub-Gaussian if   $\EP(|\xi|\geq t) \leq C e^{-c t^2}$  for any $t>0$ and some constants  $C, c>0$,  and  its  scale     is defined as $\sigma( \xi) =
\inf \{\sigma>0: \EE\{\exp(\xi^2/\sigma^2)\} \leq 2\}$. Gaussian random variables and bounded random variables  are particular instances. More generally,
  $\bsbxi\in \mathbb R^p$ is   a sub-Gaussian random vector with its scale  bounded by $\sigma$, if   $\langle \bsbxi, \bsba \rangle$ is sub-Gaussian and  $\sigma( \langle \bsbxi, \bsba \rangle )\leq \sigma \|\bsba \|_2$  for any $ \bsba\in \mathbb R^{p}$.   \ 

 The  function  proposed as the     model complexity penalty is
\begin{align}
P_o(\bsbB) = \sigma^2 [ \{q\wedge  J(\bsbB) +m - r(\bsbB)\} r(\bsbB) + J(\bsbB)\log \{e p / J(\bsbB)\}], \label{p0def}
\end{align}
where    $q=r(\bsbX)$ and $q\wedge J(\bsbB) =\min\{q, J(\bsbB)\}$.


\begin{theorem}\label{thm:oracle}
Assume  that the vectorized noise matrix, or    $\vect(\bsbE)$, is {sub-Gaussian}  with mean zero and scale bounded by $\sigma$.  Let $\hat \bsbB \in \arg \min_\bsbB \frac{1}{2}\| \bsbY - \bsbX \bsbB\|_F^2 + A P_o(\bsbB)$, where $A$ is a constant.
Then for all  sufficiently large values of  $A$,  $\hat \bsbB$ satisfies the following  oracle inequality
\begin{align}
 \EE  \left[ \max\{\| \bsbX \hat \bsbB - \bsbX \bsbB^*\|_F^2, P_o(\hat\bsbB)\}\right]    \lesssim
   \inf_{\bsbB\in \mathbb R^{p\times m}}\{ \| \bsbX \bsbB - \bsbX \bsbB^*\|_F^2 + P_o(\bsbB)\} + \sigma^2. \label{picoracle}
\end{align}
\end{theorem}


 Theorem \ref{thm:oracle} is a strong non-asymptotic result because the  obtained error rate  is uniformly better than  those by selection or rank reduction  as mentioned in  Section \ref{subsec:oracleineq}.
 Indeed, we can show that     $P_o$   gives the minimax optimal error rate in this jointly sparse setting. Moreover, \eqref{picoracle}   holds under no restrictions on $\bsbX$ or $\bsbB^*$,  and its right-hand side  takes the  infimum over all   reference signals $\bsbB\in \mathbb R^{p\times m}$. 




  The theorem  gives rise to  a  model comparison criterion. By the same reasoning as in its proof,  for any collection of random  non-zero matrices $\bsbB_1 ,  \ldots ,  \bsbB_l ,  \ldots$, if we choose the  optimal one, $\bsbB_o$,  by minimizing the following predictive information  criterion over all   given matrices 
\begin{align}
\|\bsbY - \bsbX \bsbB\|^2_F + A  P_o(\bsbB), \label{picraw}
\end{align}
then    $\bsbB_o$
satisfies
$
\EE (\|  \bsbX \bsbB_o - \bsbX \bsbB^* \|^2_F )
\leq C \inf_{l\ge 1}      \EE \{ \|  \bsbX \bsbB_l - \bsbX \bsbB^* \|_F^2 + P_o(\bsbB_l) \}
$.
    Interestingly,    $P_o$  indicates that $J(\bsbB)\log \{e p / J(\bsbB)\}$, the inflation term due to selection,    should be additive to   the degrees-of-freedom term. This is     legitimate for  sub-Gaussian noise contamination, but  to our knowledge new   when compared with     other information criteria that take  the form   of loss $ +$ $  c(n,p) \times $  degrees-of-freedom.  For example, the extended Bayesian information criterion \citep{chenchen}, derived under   $p=O(n^\kappa)$ with $\kappa>0$ and some other regularity conditions,  has a multiplicative factor  $\log n + \log p$   on  the   degrees-of-freedom of the model.
For single-response models with $m=1$,   $P_o$ simplifies to $\sigma^2[q \wedge J(\bsbB)+J(\bsbB) \log \{e p/J(\bsbB)\}]$,  which   essentially  corresponds to the risk inflation criterion \citep{FosterGeorge}, but is slightly finer. Our result applies to    any $n, p, m$. 
\subsection{Scale-free predictive information criterion}

The    predictive information criterion contains a  scale parameter $\sigma$. In sparse principal component analysis,   one can substitute  an estimate $\hat\sigma$ for the unknown $\sigma$, e.g.,   $\hat\sigma^2 = \mbox{med} (\|\bsbx_j\|_2^2/n)$ \citep{johnstoneLu2009}.
In    supervised learning, however, estimating the  scale parameter  could be as  hard  as estimating the   coefficients. We propose \ a {scale-free form of }  predictive information criterion 
  that can  bypass   $\sigma$.
  Again, no incoherence  assumption  is made for the predictor matrix.

\begin{theorem} \label{thm:sf-pic}
Let $\bsbE$ have independent and identically distributed      $ \mathcal N(0, \sigma^2)$ entries. Suppose that   the true model is parsimonious in the sense that $P_o(\bsbB^*)/ \sigma^2 < mn / A_0$ for some  constant $A_0>0$.
Consider the     criterion
\begin{align}
\|\bsbY - \bsbX \bsbB\|_F^2/\{mn - A P_o(\bsbB)/\sigma^2\},\label{eq:sf-pic}
\end{align}
where the constant   $A$ satisfies   $0<A<A_0$. Then, for sufficiently large  values of $A_0$ and $A$, any  $\hat\bsbB$ that minimizes \eqref{eq:sf-pic} subject to $P_o(\bsbB)/\sigma^2 < mn/A$   satisfies $\| \bsbX \hat \bsbB - \bsbX \bsbB^*\|_F^2\lesssim P_o(\bsbB^*)$,  with probability at least $1 -C p^{-c}-C'\exp(-c' m n )$ for some  constants $C, C', c, c'>0$.
\end{theorem}

The real  model complexity penalty   is of the form   $A_1 \times\mbox{degrees-of-freedom} + A_2 \times \mbox{inflation}$, with     constants $A_1$, $A_2$ that can be determined by  computer experiments. Experience shows that   when $\sigma^2$ is known or can be well-estimated, the choice  $A_1= 2.4$,  $A_2=1.8$ works well in \eqref{picraw}, and we recommend $A_1=2$, $A_2=1.8$ for the scale-free  form \eqref{eq:sf-pic}.
\section{Computation}
\label{sec:comp}
\subsection{A computational framework}
\label{subsec:comp}
To  ensure that selective reduced rank regression   can be applied, we must address some  challenges in  computation.   
First, the rank constraint makes    problem  \eqref{sel-rrr}   nonconvex and non-smooth. Moreover, in view of Theorem \ref{thm:oracleerr},   
      to relax  the  incoherence conditions  required by   $\ell_1$-type  penalties, nonconvex penalties  may be of   interest   \citep{mplus,ZZconcave}.   Since   different nonconvex penalty forms  may lead to   the same thresholding rule,  we study   thresholding-induced penalties.

\begin{definition}[Threshold function]\label{def:threshold}
A threshold function is a real-valued
function  $\Theta(s;\lambda)$ defined for $-\infty<s<\infty$
with $\lambda\ge 0$ as the parameter   such that
(i) $\Theta(-s;\lambda)= -\Theta(s;\lambda)$, (ii)  $\Theta(s;\lambda)\le \Theta(s';\lambda)$ for $s\le s'$,
(iii) $\lim_{s\to\infty} \Theta(s;\lambda)=\infty$, and
(iv) $0\le \Theta(s;\lambda)\le s$  for  $0\le s<\infty$.
Moreover,
 $\vec\Theta$ is defined to be  a multivariate function associated with  $\Theta$ if for any vector $\bsb{a}\in \mathbb R^{m}$, $\vec\Theta(\bsb{a};\lambda)=
\bsb{a} \Theta(\|\bsb{a}\|_2;\lambda)/\|\bsb{a}\|_2$ for $\bsb{a}\neq  \bsb{0}$ and $\bsb{0}$ otherwise. For any matrix $\bsbA\in {\mathbb R}^{p\times m}$ with $\bsbA=( \bsb{a}_1 \  \ldots \ \bsb{a}_p  )^{\tran}$, $\vec\Theta(\bsbA;\lambda) = \{ \vec\Theta({\bsb{a}}_1;\lambda) \ \ldots \ \vec\Theta({\bsb{a}}_p;\lambda)  \}^{\tran}$.
\end{definition}


Some thresholding functions, such as the hard-thresholding $\Theta(s;\lambda)=s 1_{|s|\le \lambda}$ or $s 1_{|s|< \lambda}$,   have discontinuities. To avoid   ambiguity in definition,    when using such thresholdings, we    assume that the quantity to be thresholded does not correspond to a discontinuity point.
 Let us consider the following scaled version of  problem \eqref{sel-rrr},
\begin{align}
\min_{\bsbB=(\bsb{b}_1,\ldots, \bsb{b}_p)^\tran\in \mathbb R^{p\times m}} F(\bsbB; \lambda) = \frac{1}{2K}\|\bsbY - \bsbX \bsbB\|_F^2 + \sum_{j=1}^p P(\|\bsb{b}_j\|_2;\lambda), \mbox{ s.t. } r(\bsbB)\leq r, \label{compobj}
\end{align}
where   $P$ is associated with $\Theta$ through \eqref{theta-p-coupling}  and    $K$ is a large enough number to be specified   in Theorem \ref{thm:algconv}.
 To get rid of  the low-rank constraint, we may write  $\bsbB=\bsbS \bsbV^{\tran}$,  with     $\bsbS=(\bsbs_1, \ldots, \bsbs_p)^{\tran}\in \mathbb R^{p\times r}$ and $\bsbV \in \mathbb O^{m\times r} =\{\bsbV\in \mathbb R^{m\times r}: \allowbreak\bsbV^{\tran} \bsbV=\bsbI\}$. The optimization is   now with respect to $\bsbV$ and $\bsbs_j$ ($j=1,\ldots,p$).   We   abuse notation    and    use      $F(\bsbs_1, \ldots, \bsbs_p, \bsbV;\lambda)$ to denote the objective function.
Algorithm \ref{alg:selrrr} is developed based on a block coordinate descent  method, where the   $\bsbV$-optimization can be solved by Procrustes rotation and the $\bsbs_j$  can be obtained by iterative thresholding.
\begin{algorithm}[h]
\caption{{Selective reduced rank regression   } \label{alg:selrrr}}
\small{
Data:{ $1\leq r \leq  p $, $\lambda \geq 0$, 
$\bsbS^{(0)} \in \mathbb{R}^{p\times r}$,  $\Theta$:  thresholding rule, $M_{\mbox{\footnotesize inner}}$: maximum number of inner iterations, $M_{\mbox{\footnotesize outer}}$: maximum number of outer iterations}

 1)  $t\gets 0$,  $K\gets  \|\bsbX\|_2^2$ \;

repeat

$\quad$ 2) $t\gets t+1$\;

$\quad$ 3)   Let  $\bsbW \leftarrow \bsbY^{\tran} \bsbX \bsbS^{(t-1)}$, and compute its reduced  singular-value decomposition     $\bsbW= \bsbU_w \bsbD_w \bsbV_w^{\tran}$\;

$\quad$ 4) $\bsbV^{(t)} \gets  \bsbU_w \bsbV_w^{\tran}$\;

$\quad$  5) Execute  the subroutine below   to update $\bsbS$

$\quad\qquad$  5.a)  $l \gets 0$, $\tilde \bsbS^{(0)}\gets \bsbS^{(t-1)}$\;

$\quad\qquad$ repeat

$\quad\qquad\qquad$ 5.b)  $l\gets l+1$\;

$\quad\qquad\qquad$ 5.c)  $\bsbXi^{(l,t)}\gets \bsbX^{\tran}  \bsbY \bsbV^{(t-1)}/K+ (\bsbI -  \bsbX^{\tran} \bsbX/K) \tilde \bsbS^{(l-1)}$\;

$\quad\qquad\qquad$       5.d)  $\tilde \bsbS^{(l)} \gets \vec\Theta(\bsbXi^{(l,t)}; {\lambda} )$\;

$\quad\qquad$ until {$l\ge M_{\mbox{\footnotesize inner}}$ or $\|\tilde \bsbS^{(l)}-  \tilde \bsbS^{(l-1)}\|$  is sufficiently small\;}

$\quad$  6)  $\bsbS^{(t)} \gets \tilde \bsbS^{(l)}$\;
$\quad$  7) $\bsbB^{(t)} \gets   \bsbS^{(t)} ( \bsbV^{(t)})^{\tran}$\; 
}
until {$t\ge M_{\mbox{\footnotesize outer}}$ and $\|\bsbB^{(t)}-  \bsbB^{(t-1)}\|$ is  sufficiently small. }

Output $\hat \bsbB= \bsbB^{(t)}$,  $\hat \bsbV= \bsbV^{(t)}$.
\end{algorithm}

We will show that, given any  $\Theta$,  the
algorithm  is guaranteed to    converge   under    a universal choice of $K$. 
For simplicity,   in the following theorem  we      assume that            $\vec \Theta(\cdot;\lambda)$  is continuous at any point in the closure of   $\{\bsbXi^{(l,t)}: l\ge 1, t\ge1\}$. 
The condition holds   for all continuous thresholding rules.  Practically used thresholding rules  have  few discontinuity points and such discontinuities rarely occur in  real data analysis.

\begin{theorem} \label{thm:algconv}
Given an arbitrary thresholding function    $\Theta(\cdot;\lambda)$, let    $P$ be an associated  penalty satisfying
\begin{align}
P(\theta;\lambda)-P(0;\lambda)=
\int_0^{|\theta|} [\sup\{s:\Theta(s;\lambda)\leq u\} - u] \rd u
+ {Q}(\theta; \lambda), \label{theta-p-coupling}
\end{align}
for some  ${Q}$  satisfying ${Q}(\cdot,\lambda)\ge 0$,   and  ${Q}(\theta;\lambda)=0$ if $\theta=\Theta(s;\lambda)$ for some $s\in \mathbb R$.
 Let  $K\geq \|\bsbX\|_2^2$. Then   given any starting point $\bsbS^{(0)}  \in   \mathbb{R}^{p\times r}$, 
 $F(\bsbB^{(t)};\lambda)$ converges,  $r(\bsbB^{(t)})\leq r$, and
\begin{align}
F(\bsbB^{(t)})-F(\bsbB^{(t+1)})\geq (1-\|\bsbX\|_2^2/K)\|\bsbS^{(t)}-\bsbS^{(t+1)}\|_F^2/2. \label{funcdec}
\end{align}
Furthermore, if $K>\|\bsbX\|_2^2$, then  any accumulation point  of $(\bsbs_1^{(t)}, \ldots, \bsbs_p^{(t)}, \bsbV^{(t)})$ is a  coordinatewise minimum point of $F$  and
the function value converges monotonically to $F(\bsbs_1^{*}, \ldots, \bsbs_p^{*}, \bsbV^{*})$ for some  coordinatewise minimum point $(\bsbs_1^{*},\ldots, \bsbs_p^*, \allowbreak \bsbV^{*})$.
\end{theorem}

Equation \eqref{theta-p-coupling}   covers all   aforementioned    convex and nonconvex penalties; see \cite{SheGLMTISP} for more examples.
For penalties with $Q(\cdot;\lambda) =0$, Theorem \ref{thm:algconv}  provides a stationary point guarantee.
When $\Theta$ has discontinuities,   ${Q}$  can   have infinitely many choices, which means  that different  penalties may be associated with the same thresholding function.
For instance, define  a {hard-ridge} thresholding rule   
\begin{eqnarray}
\Theta_{\mbox{{\footnotesize HR}}}(s;\lambda,\eta)=\begin{cases} 0, & \mbox{   } |s|< \lambda\\ {s}/({1+\eta}), & \mbox{  }  |s|\geq\lambda. \end{cases} \label{hybridthfunc}
\end{eqnarray}
Then,  with a nontrivial $Q$ defined by    $Q(\theta; \lambda, \eta)=0.5
  (1+\eta)\allowbreak (\lambda-|\theta|)^2 \allowbreak 1_{ 0 < |\theta| < \lambda}$, \eqref{theta-p-coupling} gives    an $\ell_0+\ell_2$  penalty $P(\theta;\lambda, \eta)= \eta \theta^2/2  + {\lambda^2} 1_{\theta\neq0} /(2+2\eta)$, or
$
P(\bsbB; \lambda, \eta)= \eta \|\bsbB\|_F^2/2  + {\lambda^2}\| \bsbB\|_{2,0}/(2+2\eta)
$ 
  in the context of  \eqref{sel-rrr}. The  Frobenius component in the hybrid penalty can shrink  the coefficients  to compensate for collinearity and large noise   in large-$p$ applications. Section \ref{subsec:screening} makes use of  a constraint variant of the penalty  for screening. 

When we apply  a component-wise  $\Theta$ in place of   $\vec\Theta$ in Step 5d, a   result similar to Theorem \ref{thm:algconv} can be obtained for the   objective function
\begin{align}
 \min_{\bsbS=(s_{j,k})\in \mathbb R^{p\times r}, \bsbV\in \mathbb O^{m\times r}}  \|\bsbY - \bsbX \bsbS \bsbV^{\tran}\|_F^2/(2K) + \sum_{j=1}^p\sum_{  k=1}^r P(|s_{j, k}|;\lambda^e).\label{sparrr}
\end{align}
The 
  sparsity is imposed on $\bsbS$ rather than on   the overall coefficient matrix $\bsbS\bsbV^\tran$.    We call \eqref{sparrr}    sparse reduced rank regression.  With  $\bsbS=(\tilde\bsbs_1, \ldots, \tilde\bsbs_r)$ and $\bsbV=(\bsbv_1, \ldots, \bsbv_r)$,  we see that    $
\bsbX \bsbS \bsbV^{\tran}$ is a sum  of $r$ factors, $\bsbX \tilde\bsbs_1 \bsbv_1^{\tran} + \cdots +  \bsbX \tilde\bsbs_r \bsbv_r^{\tran}
$, and every $\tilde \bsbs_k$ is sparse ($k=1,\ldots, r$).
Algorithm \ref{alg:selrrr} is simple to implement and has low computational complexity. When $r>1$, in addition to       some elementary matrix multiplication and  thresholding operations,   a  singular value decomposition  is carried out on $\bsbW$, which, however,     has only   $r$ columns.
To initialize the algorithm, we can   use    the reduced rank regression estimate $ (\bsbX^{\tran}\bsbX)^{+} \bsbX^{\tran} \bsbY \bsbV_r  \bsbV_r^{\tran}$ and  set   $\bsbS^{(0)}=(\bsbX^{\tran}\bsbX)^{+} \bsbX^{\tran} \bsbY \bsbV_r$,   where $^+$ denotes the Moore--Penrose inverse and $\bsbV_r$ is formed by the first $r$ eigenvectors of $\bsbY^{\tran}\bsbX (\bsbX^{\tran}\bsbX)^{+} \bsbX^{\tran} \bsbY$. Other initialization schemes are possible; see    \cite{rousseeuw1999fast}.

  From Algorithm \ref{alg:selrrr}, or  the proof of Theorem \ref{thm:algconv} in the Appendices, the optimal $\bsbV$ can be expressed in terms of $\bsbS$, i.e.,   $\bsbV_{o} (\bsbS) \allowbreak  = \allowbreak \bsbU_w \bsbV_w^\tran\allowbreak =\allowbreak \{  (\bsbX \bsbS)^\tran   \bsbY \bsbY^\tran  \bsbX \bsbS\}^{-1/2}\allowbreak (\bsbX \bsbS)^\tran \bsbY$. Hence
  $\|Y - X S V_{o}^\tran(\bsbS)\|_F^2=  \|[\bsbI - \bsbX \bsbS \{\bsbS^\tran \bsbX^\tran  \bsbY \bsbY^\tran  \bsbX \bsbS\}^{-1/2} (\bsbX \bsbS)^\tran]  \allowbreak \bsbY\|_F^2$ or $\| \bsbX \bsbS \|_F^2  - 2 \| \bsbY^\tran \bsbX \bsbS\|_*+ \|\bsbY\|_F^2$, where $\|\cdot\|_*$ is the nuclear norm. This means  that the  loading matrix obtained from \eqref{compobj} or \eqref{sparrr}    depends   on $\bsbY$ through  $\bsbY \bsbY^\tran$.

Some recent theoretical studies  \citep{BerthetR13,Gao14}  show that computationally efficient algorithms,  such as those  with polynomial time complexity, may possess an intrinsic  lower  bound in statistical accuracy that is larger than the minimax error    rate derived for most   challenging problems. This seems to  hold in our problem as well. We  will   not further pursue   this  in the current paper.


\subsection{Rank-constrained variable screening}
\label{subsec:screening}
Statisticians  are   frequently confronted with   challenges in large-scale  computing, so variable screening  has   become a popular   practice in   high-dimensional data analysis.   In multivariate problems, we are interested in     rank-constrained variable screening, which can be   achieved by the  following  form of    selective reduced rank regression
\begin{align}
\min_{\bsbB\in \mathbb R^{p\times m}} F(\bsbB) =\frac{1}{2K} \|\bsbY - \bsbX \bsbB\|_F^2  + \frac{\eta}{2} \|\bsbB\|_F^2 \, \mbox{ subject to } \|\bsbB\|_{2,0} \leq d, r(\bsbB) \leq r.  \label{selrrr-obj-constr}
\end{align}
Similar to the rank constraint, which limits the number of   factors, the  cardinality constraint, rather than a penalty, enables one to directly control the number of   predictors  selected for factor construction.  The upper bound    $d$ can be loose for the purpose of
screening, provided it is not too small.

We show below how to   use     a {quantile} version of the hard-ridge thresholding \eqref{hybridthfunc} to solve such problems.
Given $1\le {d} \le   p$,  $\eta\ge 0$, for any $\bsb{s}=(s_1, \ldots, s_p)^{\tran}\in {\mathbb R}^p$,   $\Theta^{\#}(\bsb{s}; {d}, \eta)$ is  defined to be a vector $\bsb{t}=(t_1, \ldots, t_p)^{\tran} \in {\mathbb R}^p$ satisfying $t_{(j)} = s_{(j)}/(1+\eta)$ if $1\leq j \leq {d}$, and $0$ otherwise. Here,  $s_{(1)}, \ldots, s_{(p)}$ are the order statistics of $s_1, \ldots, s_p$, i.e.,  $|s_{(1)} | \ge \cdots \ge |s_{(p)}|$, and $t_{(1)}, \ldots, t_{(p)}$ are defined similarly. In the case of ties, a random tie breaking rule is used.
The multivariate quantile thresholding function    $\vec{\Theta}^\# (\bsbS; {d}, \eta)$  for any $\bsbS=(\bsbs_1, \ldots, \bsbs_p)^{\tran}\in \mathbb{R}^{p\times r}$, is defined as a $p\times r$ matrix ${\bsbT}=(\bsb{t}_1, \ldots, \bsb{t}_p)^{\tran}$ with ${\bsb{t}}_j=\bsbs_j/(1+\eta)$ if $\|\bsbs_j\|_2$ is among the ${d}$ largest  elements in   $\{\|\bsbs_j\|_2: 1\leq s \leq p\}$,  and $\bsb{0}$ otherwise.
%
%
Now we modify Step 5.d of Algorithm \ref{alg:selrrr} to         $\tilde \bsbS^{(l)} \gets \vec\Theta^{\#}(\bsbXi^{(l,t)}; d, \eta )$, and   all other steps remain unchanged. The resulting    algorithm for rank-constrained screening always converges.

\begin{theorem}\label{thm:algconv-scr}
Assume $K \geq \|\bsbX\|_2^2$. 
  Then, given any $\bsbS^{(0)}\in \mathbb R^{p\times r}$, $F(\bsbB^{(t)})$  is non-increasing and satisfies  $F(\bsbB^{(t)})- F(\bsbB^{(t+1)}) \geq (1-\|\bsbX\|_2^2/K)\|\bsbS^{(t)}-\bsbS^{(t+1)}\|_F^2/2$, and   $\bsbB^{(t)}$ obeys the constraints $\|\bsbB^{(t)}\|_{2,0} \leq {d}$ and $r(\bsbB^{(t)})\leq r$   for any $t\geq 1$.
\end{theorem}

To get some intuition, let us set   $\bsbS^{(0)}=\bsb{0}$. Then,   at the first iteration, $\bsbW=\bsb{0}$, $\bsbV^{(1)}=\bsbI$, and the quantile thresholding      picks     $d$ features according to  the   marginal statistics  $\bsbX^{\tran} \bsbY$, which  amounts to      sure independence screening  \citep{fanlv}.  Our algorithm 
iterates further to lessen the  greediness of independence screening.
To accelerate the computation, we    recommend progressive screening  in the iterative process. Concretely,   we use  a  sequence    $Q(t)$ that decreases from $p$ to ${d}$, e.g., $Q(t)=
{2 p}/\{  {1+\exp(\alpha t)}\}$ with $\alpha=0.01$ and $0\le t \le  (1/\alpha) \log(2p/d-1)$, and perform  $\tilde \bsbS^{(l)} \gets \vec\Theta^{\#}(\bsbXi^{(l,t)}; Q (t), \eta )$ in Step  5d;         after obtaining $\bsbB^{(t)}$ in Step 7,     the following data  squeezing operations are carried out,
$
 \mathcal J\gets \{j: \bsbS^{(t)}(j,1:r)\neq \bsb{0}\}$,
$\bsbS^{(t)} \gets \bsbS^{(t)}(\mathcal J, 1:r)$,    $  \bsbX \gets \bsbX(1:n, \mathcal  J)$. 
  An attractive feature of the implementation is that
  as the cycles progress,     the problem size  drops   quickly and  the      computational load can be significantly reduced.

For the sparse reduced rank regression   with an   $\ell_0$-{constraint}
\begin{align}
\min_{\bsbS\in\mathbb R^{p\times r}, \bsbV\in \mathbb O^{m\times r}} \frac{1}{2K} \|\bsbY - \bsbX \bsbS\bsbV^{\tran}\|_F^2 +  \frac{\eta}{2} \|\bsbS\|_F^2  \mbox{ subject to } \|\bsbS\|_{0} \leq  {d}^e, \label{sparrr-obj-constr}
\end{align}
   similar algorithms can be developed based  iterative quantile thresholding.  
In big data applications, a good idea is to  combine \eqref{selrrr-obj-constr} with   \eqref{sparrr-obj-constr}, because  calling the rank-constrained  screening algorithm    in an earlier stage can reduce the dimensionality from $p$ to $d$.  In this hybrid scheme,    $d$ satisfies   ${d}\leq {d}^e\leq  {d} r$, and          ${d}={d}^e$ gives a conservative screening choice.  


\section{Unsupervised   Selective and Sparse Principal Component Analyses}
\label{sec:unsup}

This section studies  selective factor construction in   principal component analysis. We assume that only one  data matrix  $\bsbX\in \mathbb R^{n\times p}$   is available and it   has been column-centered.
Principal component analysis   can be interpreted as finding  a low-rank matrix      $\bsbB$ to approximate the observed data. Similar to Section \ref{subsec:comp}, we write    $ \bsbB= \bsbV \bsbS^{\tran}$ with   $\bsbS = (s_{j,k}) = ( \bsbs_1, \ldots, \bsbs_p)^{\tran}\in \mathbb R^{p\times r}$ and $\bsbV\in \mathbb O^{n\times r}$. The selective principal component analysis problem is defined as 
\begin{align}
\min_{\bsbS \in \mathbb R^{p\times r}, \bsbV \in \mathbb O^{n\times r}} \frac{1}{2} \| \bsbX - \bsbV  \bsbS^{\tran} \|_F^2 + \sum_{j=1}^p P(\|\bsb{s}_j\|_2; \lambda).  \label{sel-pca}
\end{align}
  Obviously,  \eqref{sel-pca} can be  rephrased
as a special case  of    selective reduced rank regression, by taking  $\bsbX^{\tran}$ as the response matrix and $\bsbI_{p\times p}$ as the design matrix.
Likewise, adapting    \eqref{sparrr} to     the unsupervised setting leads to  the following criterion for       sparse principal component analysis,
\begin{align}
\min_{\bsbS\in \mathbb R^{p\times r}, \bsbV\in \mathbb O^{p \times r}}  \frac{1}{2}\|\bsbX -   \bsbV\bsbS^{\tran} \|_F^2 + \sum_{j=1}^p\sum_{  k=1}^r P(|s_{j, k}|;\lambda^e).\label{spapca}
\end{align}
 Unsupervised versions of   \eqref{selrrr-obj-constr} and \eqref{sparrr-obj-constr}\  can also be defined. Moreover, based on the discussions in    Section \ref{subsec:comp}, all these criteria   depend on $\bsbX$ through $\bsbX^\tran \bsbX$. 

 Problems \eqref{sel-pca} and     \eqref{spapca}   are    perhaps less challenging  than their supervised counterparts, but    they still     provide   new insights into sparse principal component analysis. For example,      \eqref{spapca}     defines a multivariate criterion that is able to find  all sparse loading vectors simultaneously.
In computation, our algorithms developed in Section \ref{sec:comp}    simplify greatly. In fact, because of the identity design, the inner loop in Step 5  of Algorithm \ref{alg:selrrr}   converges in one iteration     and  the overall procedure reduces to
\begin{align}
\bsbX \bsbS^{(t-1)}  = \bsbU \bsbD \bsbV^{\tran} ,     \quad \bsbS^{(t)}  \gets \vec\Theta (  \bsbX^{\tran} \bsbU \bsbV^{\tran}; {\lambda}).\label{reducedselpcaalg}
 \end{align}
 In other words,  $\bsbS^{(t)}$ is updated  by thresholding    $\bsbX^\tran \bsbX \bsbS^{(t-1)} \{( \bsbS^{(t-1)})^\tran\bsbX^\tran\bsbX \allowbreak \bsbS^{(t-1)}\}^{-1/2}$, and various thresholding operators can be used.
When $r=1$, the singular  value decomposition is unnecessary, since  $\bsbU \bsbV^{\tran}$ can be directly obtained by normalizing the column vector      $
\bsbX \bsbS^{(t-1)}$.

We now point out some   recent literature related to     \eqref{sel-pca} and \eqref{spapca}.
Under a spiked covariance model assumption, \cite{cai2013}        proposed   and studied an adaptive multi-step procedure to solve a  problem similar to  \eqref{sel-pca}. Our    algorithm  \eqref{reducedselpcaalg} is   closest in spirit to the thresholding procedure in  \cite{johnstoneLu2009}.         \cite{Ma2013} proposed an iterative algorithm  for principal subspace estimation,  but it    has no  guarantee of numerical  convergence. 
Another type of sparse principal component analysis   sets $\bsbY =\bsbX$  in      \eqref{sparrr}; the idea seems to first appear  in  \cite{spapca}.   
But the self-regression    formulation may bring some  ambiguity in selection. 
Consider  a noise-free model  where   $\mathcal J^*$   gives the set of indices of all nonzero columns in $\bsbX$, and    $r^*=r(\bsbX)$   obeys $r^* < |\mathcal J^*|$. Then, given   any  index set $\mathcal J\subset \{1, \ldots, p\}$ satisfying $r(\bsbX_{\mathcal J})=r^*$,     we can find a matrix $B$ with $  \mathcal J(\bsbB)\subset \mathcal J$ and $ r(\bsbB)\le  r^*$ such that   $\bsbX = \bsbX \bsbB$, but
   $|\mathcal J|$ can be smaller or larger than $|\mathcal J^*|$. 

\section{Applications}
\label{sec:expr}
\subsection{Paper quality data}
 \cite{Aldrin96} described two paper datasets collected from Norwegian paper industry.      We focus on the first dataset. The production of paper depends on  a huge number of predictors. The data were obtained by varying   three control variables $x_1$, $x_2$, $x_3$ that are coded as   $1$,  $0$, and $-1$.    There are   $15$ design points for $(x_1, x_2, x_3)$. At each  point the paper quality was measured twice,  except once at design point $(0, 1, 1)$,   which results in $29$ observations.  The response variables are the 13 measures  used to evaluate   paper quality.

The first model we considered is a   full quadratic model  recommended by  Aldrin, with $9$ predictors and $13$ responses in total.
Previous analysis 
suggests that  the model   is factor driven. Reduced rank regression can be used  to construct   explanatory score variables for  paper quality assessment  \citep{Aldrin96,izenbook}, but has to keep    all   predictors  in the final model. Our proposed  approach has the ability of  identifying a small set of  predictors for       selective factor extraction.
We split  the whole dataset at random, with  60\% for  training and     $40\%$  for test, and repeat the process for      $100$ times  to  compare the performance of the two methods. All  parameters were tuned  by the scale-free predictive information criterion.   Based on the   median statistics, both   reduced rank regression and selective reduced rank regression  gave two factors, and   showed comparable test errors,  $167$ and $168$, respectively. The difference is that    selective reduced rank regression  achieved  this with only   $5$ predictors, about half of the  model size of  reduced rank regression.

A careful examination of the data   shows some  interesting findings that merit further investigation.    Although sample 5 and sample 6 correspond to  the same design point $(0, 0, 0)$,  their response values   demonstrate   larger-than-normal discrepancies. Take the first response variable $\bsby_1$ for instance. We get      $y_1(5)- y_1(6)=-1.7831$, but  the differences  in $\bsby_1$ for  all the other observation pairs are   bounded by   $0.4458$ in absolute value.
 We suspect that there   exist outliers in the data. Their occurrence   might be  due to the crude  coding  of   control variables when they were varied.   Therefore, a     robust factor model  is perhaps more appropriate. Specifically,   the factors constructed from input variables,  $\bsbF=\bsbX\bsbA$, may need   correction  at certain anomaly points, which     amounts  to       $\bsbY = (\bsbF+ \bsb{O}) \bsbC +\bsbE$. The correction term    $\bsb{O}$ is desired to be    row-wise sparse, since   the outliers should
not be the norm. Therefore, we have      $$\bsbY = \bar\bsbX \bar\bsbB + \bsbE, \quad \bar\bsbX=[\bsbX \,\,  \bsbI]\in \mathbb R^{n\times (n+p)},$$
where  $\bar \bsbB$ is desired to have    low rank and sparse rows. For  such an augmented model, the number of predictor variables  is always larger than the sample size.  We repeated the analysis with   the augmented  design. In    the 100 experiments,     two   variables, $\bsbX_1$ and $\bsbX_3$, got selected all the time; the other predictors, including all interaction terms, never entered the model. The median model size is however   $5$, due to the existence of   some automatically detected outliers.   Sample $6$ and sample $7$, corresponding to design points $(0, 0, 0)$ and $(-1, 0, 1)$, respectively,  were identified as anomalies more than  $50\%$ of the time. The  rank   is now as low as $1$, but the test error is  substantially reduced to  $148$. 

\subsection{Macroeconomic  data}
\cite{SW11} summarized  194 quarterly observations on 144   macroeconomic time series   observed from 1960  to 2008, with some earlier observations used for lagged values of regressors as necessary.
We preprocessed the data using  the   transformations given  in  Table B.2 of \cite{SW11}. One variable,  non-borrowed reserves of depository institutions, was removed because its transformation involves logarithms  but it has negative   values.  Of the 143 series, 35 are high-level aggregates, the information of which is all contained in the rest. 
Our    predictors are   the 108 disaggregated series and their lagged values.
The series are grouped into thirteen broad categories. We use  the interest rates category, which  consists of 13 time series,  as our response variables.
The  dataset is a good example to show that      although forecasters and policymakers can access many  potentially relevant macroeconometric time series,   excluding noninformative ones is often  {ad hoc}. 
In fact, to the best of our knowledge, there is no acknowledged model  to  describe all the 13 interest rates  covering treasuries, corporate, term spreads and public-provide spreads.
Low-rank models naturally arise   \citep{ReinVelu}, and the necessary factors can be as few as two or even one, which has been theoretically established in  a large body of  economics literature.

First, we used the 108 series observed in the past four quarters as predictors, giving       $p=4\times 108=432$ predictors  and   $432\times 13$ unknowns.
We used Algorithm \ref{alg:selrrr} for     selective reduced rank regression  and    the scale-free predictive information criterion for parameter tuning. The resulting       model has   $\hat r = 1$ and $\hat J=3$, achieving  a remarkable  dimension reduction. The single explanatory  factor in response to all interest rates series  is constructed from    capital utilization,    the 3-month treasury bill secondary market rate  minus the   10-year treasury constant maturity rate, and  Moody's  Baa corporate bond yield minus   the 10-year treasury constant maturity rate.
 Since   no variables of lag order 2 or above were selected, we   repeated the analysis using the series with   only one lag, thus selecting    four variables and   two factors. The last two variables shown  in the 1-factor model appear again, and       the employment category contributes the other two variables,   relating to employees on nonfarm payrolls in   wholesale trade   and   help-wanted advertising in newspapers.
Both the 1-factor model and the 2-factor model      show a high level of parsimony.

Next,  we did a forecasting experiment to compare the  obtained factor models   with     auto-regressive modeling with four lags, which  is a conventional but quite  accurate  forecasting method.
The  performance is evaluated by a  rolling scheme: a  rolling estimation window of the most recent 100 quarterly observations   is used  to estimate the parameters, and   forecasts are made in the forecast window. Both  windows      move  forward by one quarter at a time, and the procedure is repeated  94 times.
Table \ref{Tab:intratesRolling} shows the mean squared errors       of each method for  the 13 interest rates. Overall, the three methods have comparable  prediction errors. Of course, the comparison is, in some sense, unfair to factor models. The auto-regression method builds  a separate model for each  interest rate using four relevant predictors,  while the 1-factor method, say,  regresses every response on the same single score variable. Our purpose here is to demonstrate the usefulness of   category-level factors; better   models can be possibly built on the factors to improve the accuracy further.

\begin{table}[h!]
 \small
{
\caption{\emph{Mean-squared errors  of AR(4), the one-factor model, and the two-factor model, for 13 time series in the  interest rate category, with their  medians and means reported in the last column}}
\setlength{\tabcolsep}{1.1mm}
\begin{tabular}{l ccc ccc ccc cccc c}
\\
Series index&  1  & 2  & 3  & 4 & 5 & 6 & 7    & 8 & 9 & 10 & 11 & 12 & 13  & {(med,mean)}\\
AR(4)  & 9.2 & 10.6 &10.3&13.5 & 12.3 & 8.7 & 7.1      & 9.0 & 0.7 &  2.4 & 12.7  &1.7 & 1.9& {(9.0,7.7)}\\
$1$-factor & 9.7 & 7.8 & 8.4 & 8.8 & 16.0 & 8.0 & 5.4   & 6.9 & 1.7 & 10.1 & 14.2 & 8.0 & 8.8& {(8.4,8.8)} \\
$2$-factor    &  9.5 &  8.4 &  8.8  &  12.7  &  10.4 &  8.4&  5.3   &  7.3 &  1.3 &  8.1  &  11.1  &  6.4 &  8.3& {(8.3,8.1)}\\
\end{tabular}
\label{Tab:intratesRolling}}
\end{table}




\subsection{Face data}
\begin{figure*}[h!]
\centering
  \subfloat{\label{fig:faceExs}\includegraphics[width=0.75\textwidth]{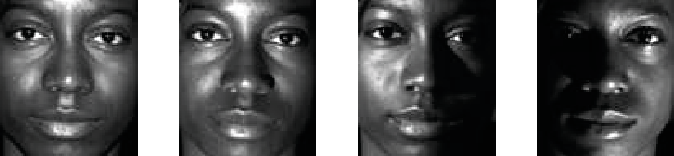}} \\
  \subfloat{\label{fig:selpcaExs}\includegraphics[width=0.75\textwidth]{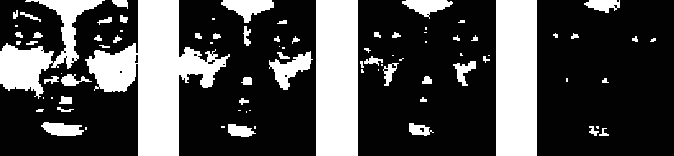}}\\
  \caption{ Upper panel:  face image examples for subject 22 with image numbers  $1, 10, 20, 50$. Lower  panel: regions of interest,  denoted by white,  identified by selective principal component analysis   with $r=30$, and  ${d}=2400, 1200, 600, 200$ from left to right.
  }
  \label{fig:yaleb-res1}
\end{figure*}

 The Extended Yale Face Database B
\citep{KCLee05,GeBeKr01}  contains  aligned and  cropped face images of 38 subjects with the same frontal pose under 64 different illumination conditions.
We down-sampled the images to $96\times 84$, each containing $8064$ pixels.   Given a subject, a data matrix of size 64$\times$8064 can be formed from the associated   images.
  In face recognition, principal component analysis is widely used  to extract      basis features, referred to as {eigenfaces}, the number of which  is controlled by the rank. We set $r=30$ throughout this experiment and focus on the 22nd subject  in the database, whose     image examples  are shown in the upper panel of Fig. \ref{fig:yaleb-res1}.
 Selective principal component analysis was performed   in   the hope of     capturing       regions of interest  under different   light source directions. As seen in the lower panel,   some informative regions  sensitive  to   illumination conditions, e.g.,      forehead and nose tip, were automatically detected.

\begin{table}[h!]
\small{
\caption{\emph{Performance comparison between  sparse and hybrid principal component analyses  in terms of computational time, adjusted variance   percentages (AV), and the number  of pixels involved}}
 \begin{tabular}{r ccc ccc ccc  }
           & \multicolumn{3}{c}{${d}^e=4800 $} & \multicolumn{3}{c}{${d}^e=2400 $} & \multicolumn{3}{c}{${d}^e=1200 $}   \\
\cmidrule(r){2-4} \cmidrule(l){5-7} \cmidrule(l){8-10}
        & Time & Pixels  & AV & Time & Pixels  & AV & Time & Pixels  & AV\\
Sparse &            254& 4435& 51.3& 278& 2370& 33.4& 293& 1198& 20.8\\
Hybrid&  131& 4377& 51.1& 87 & 2187& 35.9& 81 & 1127& 20.5 \\
\end{tabular}
\label{tab:spapcavss2pca}
}
\end{table}
%
%

 To reduce the computational burden caused by the large number of pixels,  we     tested  the     screening-guided hybrid sparse principal component analysis. See its description   below \eqref{sparrr-obj-constr}     and recall that       ${d}^e$ controls the number of nonzero elements in the loading matrix $\bsbS$, and     ${d}$ controls  the number of nonzero rows.   Table \ref{tab:spapcavss2pca} shows the results with a  conservative screening choice    ${d} = {d}^e$. The adjusted variance rates were computed according to \cite{shenhuang}. The hybrid principal component analysis gave  essentially the same  adjusted variances as   sparse principal component analysis, but used fewer pixels. More importantly, the hybrid approach   offered impressive time savings. 
Then, we   did   an experiment with  a less conservative screening choice,         ${d}=2400$,  ${d}^e=3600$. Sparse principal component analysis used 3517 pixels   to reach an adjusted variance rate of  $43\%$, while hybrid principal component analysis   reduced the model size   to $2400$ pixels, and gave adjusted variance rate    $40\%$. When using   2400 pixels,  sparse principal component analysis only reached an adjusted variance rate  of    $34\%$.

\section{Discussion}
\label{sec:discussion}
The techniques we developed to study selective reduced rank regression       are applicable to  pure variable  selection or rank reduction.  For example, the recipe for proving Theorem \ref{thm:oracleerr} can handle
Schatten $ p$-norm penalized    trace regression models, without using   the sophisticated   quasi-convex Schatten class embeddings \citep{Rohde11}.
 The scale-free   predictive information criterion      addresses the   issue  of adaptive   rank selection in $p\gg n$ models, as raised by \cite{BuneaSheWeg}. 

In this work, all the problems under consideration are  nonconvex. In common with   papers like   \cite{Rohde11} and \cite {BSW12}, we studied the properties of  global minimizers. In some less challenging situations, the proposed algorithms, when initialized by the reduced rank regression estimator,    can deliver a good estimate within   a few iteration steps. This  was also evidenced by \cite{masun14} in a recent technical report. In some  hard cases, we found the multi-start strategy of \cite{rousseeuw1999fast} to be quite effective. The study of how to initialize  and when to terminate  is beyond  the scope of the current paper, but is an interesting  topic for further research.

\appendix
\section{Proofs of Theorems}\label{app:th}
Throughout the proofs, we use $C$, $c$, $L$ to denote   constants. They are not necessarily the same at each occurrence. Given any matrix $\bsbA$, we use $CS(\bsbA)$ and $RS(\bsbA)$ to denote its column space and row space, respectively. Denote by    $\Proj_{\bsbA}$  the orthogonal projection matrix onto   $CS(\bsbA)$, i.e., $\Proj_{\bsbA}=\bsbA(\bsbA^{\tran}\bsbA)^{+}\bsbA^{\tran}$, where $^+$ stands for  the Moore-Penrose pseudoinverse, and $\Proj_{\bsbA}^{\perp}$ the projection onto  its orthogonal complement. Finally, we use $[p]$ to denote $\{1,\ldots,p\}$. 
Recall that  for $\bsbA=(\bsba_1, \ldots, \bsba_p)^\tran$, $\|\bsbA\|_{2,0}=\sum_{j=1}^p 1_{\|\bsba_{j}\|\ne \bsb0}$, $\|\bsbA\|_{2,1} = \sum_{j=1}^p  \|\bsba_j\|_2 $, $\|\bsbA\|_F=\{\mathrm{tr}(\bsbA^\tran\bsbA)\}^{1/2}$, and $\|\bsbA\|_2$ is the  the largest singular value of $\bsbA$.

\subsection{Proof of Theorem \ref{thm:oracleerr}}
\label{app:subsec:oracleerr}

Recall the following three basic penalties
  $$P_1(t; \lambda) = \lambda |t|, \quad P_0(t; \lambda) = {\lambda^2} 1_{t\neq 0}/2, \quad P_H(t; \lambda)= (-t^2/2+\lambda |t|)1_{|t|<\lambda} +(\lambda^2/2) 1_{|t|\geq \lambda}.
$$
It is easy to verify that both $P_1(t; \lambda)$ and  $P_0(t; \lambda)$ are bounded below by $P_H(t; \lambda)$.  For convenience, we use $P_{2,1} (\bsbB; \lambda)$  to denote $\lambda \| \bsbB\|_{2,1}$;  $P_{2,0}$ and $P_{2, H}$  are used similarly.  Let us consider a scaled selective reduced rank regression problem
\begin{align}
\min_{\bsbB: r(\bsbB)\leq r} \| \bsbY - \bsbX \bsbB\|_F^2/2 + \sum_{j=1}^p P(\rho\|\bsb{b}_j\|_2; \lambda),
\end{align}
where  $\rho=\|\bsbX\|_2$ and $P\geq P_H$.
We also write the penalty term as $P(\rho \bsbB; \lambda)$ when there is no ambiguity. Then for any  globally optimal solution   $\hat \bsbB$, we have the  inequality below for all $\bsbB$ satisfying $ r(\bsbB)\leq r$
\begin{align}
\frac{1}{2}\| \bsbX \hat \bsbB - \bsbX \bsbB^*\|_F^2 \leq \frac{1}{2}\| \bsbX \bsbB - \bsbX \bsbB^*\|_F^2 +   P(\rho\bsbB; {\lambda}) - P(\rho\hat \bsbB; {\lambda})\
+ \langle \bsbE, \bsbX \bsbDelta \rangle. \label{est:firstineq}
\end{align}
Here,  $\bsbDelta=\hat \bsbB -  \bsbB$ and so $r(\bsbDelta)\leq 2r$.
\begin{lemma} \label{lemma:phostochastic}
There exist universal constants $A_0, C, c>0$ such that for any $a\geq {A_0^{-1/2}}$, the following event, denoted by  $\mathcal A$,
\begin{align}
\sup_{1\le r \le m}\sup_{\bsbDelta: r(\bsbDelta)\le r} [2\langle \bsbE, \bsbX  \bsbDelta\rangle - \frac{1}{ a} \|\bsbX  \bsbDelta \|_F^2 - 2aA_0\{  P_{2,H}(\rho\bsbDelta; {\lambda^o}) + \sigma^2 r(m-r\} ] \geq a  \sigma^2 t \label{event0}
\end{align}
occurs with probability at most  $ C \exp(-c t)$ for any $t > 0$, where $\lambda^o=\sigma ({r + \log p})^{1/2}$.
\end{lemma}

  Lemma \ref{lemma:phostochastic}    is the key to bound the last stochastic term in \eqref{est:firstineq}. The proof of the lemma    is based on a combination of  statistical  and computational analyses as stated by the following two lemmas.
\begin{lemma} \label{lemma:phcomp}
Given any $\bsbX$ satisfying $ \|\bsbX\|_2\le 1$, there exists a globally optimal solution $\bsbB^o=(\bsb{b}_1^o,\ldots, \bsb{b}_p^o)^{\tran}$ to the problem
\begin{align*}
\min_{\bsbB: r(\bsbB)\leq r} \frac{1}{2}\| \bsbY - \bsbX \bsbB\|_F^2 +  P_{2,H}(\bsbB; {\lambda})
\end{align*}
such that for any $j= 1,\ldots, p$, either $\bsb{b}_j^o=\bsb{0}$ or $\|\bsb{b}_j^o\|_2 \geq \lambda$.
\end{lemma}

\begin{lemma} \label{concenGauss}
Suppose the entries of $\bsbE$ are independent and identically distributed as $ \mathcal N(0, \sigma^2)$. For any given $  1\le J \leq p, 1\leq r\leq J$, define   $R_{\mathcal J, \bsbV} = \|\Proj_{\bsbX_{\mathcal J}} \bsbE \bsbV \|_F^2= \|\Proj_{\bsbX_{\mathcal J}} \bsbE \Proj_{\bsbV} \|_F^2$ and $R_{J, r}=\sup_{|\mathcal J|=J, \bsbV\in\mathbb O^{m\times r}} R_{\mathcal J, \bsbV}$, where $\mathcal J \subset [p]$, $|\mathcal J|=J$ and $\bsbV\in \mathbb O^{m\times r}$. Let $
P_o(J, r) = \sigma^2 \{(q\wedge  J +m - r) r + J\log (e p / J)\}
$.
Then for any $t> 0$,
\begin{align}
\EP\{R_{J, r} > t \sigma^2+ L P_o(J, r)\}\leq C \exp(-c t), \label{chisqresbound}
\end{align}
where $L, C, c>0$ are universal constants.
\end{lemma}

 Define    $l_H(\bsbDelta, r) =  2\langle \bsbE, \bsbX  \bsbDelta\rangle -  \|\bsbX  \bsbDelta \|_F^2 /a - 2aA_0\{  P_{2,H}(\rho\bsbDelta; {\lambda^o}) + \sigma^2 r(m-r)\}$ and $R=\sup_{1\le r \le m}\sup_{\bsbDelta: r(\bsbDelta)\le r}  l_H(\bsbDelta, r)$.
Then for  $\bsbDelta=\hat \bsbB -  \bsbB$
\begin{align*}
2\langle \bsbE, \bsbX  \bsbDelta\rangle
 \leq  \frac{1}{ a} \|\bsbX  \bsb\bsbDelta \|_F^2 + 2aA_0\{  P_{2,H}(\rho\bsbDelta; {\lambda^o}) + 2\sigma^2 r(m-r)\} + R,
\end{align*}
where  $\EE R \le C \sigma^2$ due to Lemma \ref{lemma:phostochastic}.
Substituting the bound  into \eqref{est:firstineq}, we have
\begin{equation}
\begin{aligned}
 \|\bsbX \hat \bsbB - \bsbX \bsbB^*\|_F^2 \leq  &\|\bsbX  \bsbB - \bsbX \bsbB^*\|_F^2 +4 aA_0 \sigma^2 r(m-r) +  \frac{1}{ a} \|\bsbX  \bsb\bsbDelta \|_F^2 + R \\
 &+2  P(\rho \bsbB; {\lambda}) - 2 P(\rho \hat \bsbB; \lambda)+2 a A_0  P_{2,H}(\rho \bsbDelta; \lambda^o).
\end{aligned}\label{intermedeq}
\end{equation}

When $P=P_{2,0}$, $P_{2,0}(\rho \bsbDelta; {\lambda^o})=P_{2,0}(\bsbDelta; {\lambda^o})$ and
$$ P_{2,H}(\rho\bsbDelta; {\lambda^o}) \leq  P_{2,0}(\rho \bsbDelta; {\lambda^o}) \leq  P_{2,0}(\bsbB; {\lambda^o}) +  P_{2,0}(\hat\bsbB; {\lambda^o}).$$
By the Cauchy-Schwartz inequality,   $\|\bsbX  \bsb\bsbDelta \|_F^2/a\le 2\|\bsbX  \bsbB - \bsbX \bsbB^*\|_F^2/a + 2\|\bsbX  \hat\bsbB - \bsbX \bsbB^*\|_F^2/a  $.
The oracle bound thus follows, by choosing  $a> 2\vee  { A_0^{-1/2}} $ and $\lambda = A \lambda^o$ with $A> ({aA_0})^{1/2}$.

When $P=P_{2,1}$,   we assume the condition
\begin{align}
(1+\vartheta) \rho \| \bsbDelta_{\mathcal J}\|_{2,1} \le K   J^{1/2} \| \bsbX \bsbDelta\|_F +  \rho \| \bsbDelta_{\mathcal J^c}\|_{2,1}, \label{regcondinth1}
\end{align}
where    $\mathcal J= \mathcal J(\bsbB)$ and   $J=J(\bsbB)$.
 It follows that for $\theta = \vartheta/(1+\vartheta)$,
\begin{equation}
\begin{aligned}
&(1+\theta) \rho\lambda \| \bsbDelta_{\mathcal J}\|_{2,1}
\\ \le & (1-\theta) K   J^{1/2}\lambda \| \bsbX \bsbDelta\|_F +  (1-\theta)\rho\lambda \| \bsbDelta_{\mathcal J^c}\|_{2,1}\\
 \le &(1-\delta)\| \bsbX \bsbDelta\|_F^2/2 + \{(1-\theta)^2/(2-2\delta)\} K^2\lambda^2 J +  (1-\theta)\rho \lambda \| \bsbDelta_{\mathcal J^c}\|_{2,1}, \end{aligned}\label{l1eq1}
\end{equation}
 where   $\delta$ can be any number in $(0,1)$, e.g., $\delta=0.5$.

Under $\lambda=A\lambda^o$ with $A\ge \theta a A_0$,
 we have the following result from the sub-additivity of $P_1$
\begin{equation}
\begin{aligned}
&2 \rho \lambda \|\bsbB\|_{2,1} - 2\rho \lambda \|\hat \bsbB\|_{2,1}+2 a A_0 P_{2,H}(\rho \bsbDelta; \lambda^o)\\
\le \ & 2\rho \lambda \|\bsbDelta_{\mathcal J}\|_{2,1} - 2\rho \lambda \|\bsbDelta_{\mathcal J^c}\|_{2,1} + 2 a A_0 P_{2,1}(\rho \bsbDelta; \lambda^o) \\
\le \ &  2 (1+\theta) \rho \lambda \|\bsbDelta_{\mathcal J}\|_{2,1} - 2(1-\theta)\rho \lambda \|\bsbDelta_{\mathcal J^c}\|_{2,1}.
\end{aligned}\label{l1eq2}
\end{equation}
Adding \eqref{intermedeq}, \eqref{l1eq1} and \eqref{l1eq2} together, we get
\begin{align*}
 \|\bsbX \hat \bsbB - \bsbX \bsbB^*\|_F^2 \leq &  \|\bsbX  \bsbB - \bsbX \bsbB^*\|_F^2 +4 aA_0 \sigma^2 r(m-r) + (1+ \frac{1}{ a} -\delta)\|\bsbX  \bsb\bsbDelta \|_F^2  \\ &
+\{(1-\theta)^2/(2-2\delta)\} K^2\lambda^2 J +R.
\end{align*}
Choosing $a >  \delta^{-1} \vee  A_0 ^{-1/2}$ and $A\ge \theta a A_0$ gives the desired oracle inequality.

{\remark
A popularly used regularity condition in the studies of  $\ell_1$-type penalties is the restricted eigenvalue assumption; see  \cite{bickel09},  \cite{lounici-2010}  and   \cite{BSW12}. We   extend it to  our setting as follows. Specifically,
  $\bsbX\in {\mathbb R}^{n\times p}$ satisfies  the restricted eigenvalue condition  for an index set $\mathcal J\subset[p]$  and  positive number   $\kappa$, if and only if
\begin{eqnarray}\label{condition1}
\|\bsbX \bsbDelta\|_F^2   \ge \kappa \|\bsbX\|_2^2  \sum_{ j \in \mathcal J}  \|\bsbdelta_j\|_2^2 \label{REwithnorm}
 \end{eqnarray}
holds for all  $\bsbDelta=[\bsbdelta_1, \ldots,  \bsbdelta_p ]^\tran\in\mathbb R^{p\times m}$       satisfying  $$ (1+\vartheta)\sum_{ j \in\mathcal J}  \|\bsbdelta_j\|_2  \geq  \sum_{ j \in {\mathcal J}^c}  \|\bsbdelta_j\|_2,$$  where $\vartheta>0$ is a constant.
The version defined by  \cite{BSW12}, after correcting the typo,   is slightly more stringent than \eqref{REwithnorm}.

In comparison with the multivariate restricted eigenvalue assumption, our regularity condition \eqref{regcondinth1}   has a simpler form and is less demanding. To show this,   consider two cases. The case $(1+\vartheta) \| \bsbDelta_{\mathcal J}\|_{2,1} \le \| \bsbDelta_{\mathcal J^c}\|_{2,1} $ is trivial. Suppose the reverse inequality holds. Then  $$(1+\vartheta) \| \bsbX\|_2 \| \bsbDelta_{\mathcal J}\|_{2,1} \le (1+\vartheta) J^{1/2} \| \bsbX\|_2\| \bsbDelta_{\mathcal J}\|_F  \le (1+\vartheta) J^{1/2} \| \bsbX\bsbDelta\|_F/\kappa^{1/2},$$ and so \eqref{regcondinth1} holds with  $K=(1+\vartheta)/\kappa^{1/2}$.
}

\subsection{Proof of Theorem \ref{thm:oracle}}
\label{appsubsec2}


In this proof, given any index set  $\mathcal J\subset [p]$, $\Proj_{\bsbX_{\mathcal J}}$  is abbreviated to  $\Proj_{\mathcal J}$    when there is no ambiguity.
%
%
Because $P_o(\bsbB) = \sigma^2 [\{q\wedge  J(\bsbB) +m - r(\bsbB)\} r(\bsbB) + J(\bsbB)\log \{e p / J(\bsbB)\}]$ only depends on $J(\bsbB)$ and $r(\bsbB)$, we also denote it  by  $P_o\{J(\bsbB), r(\bsbB)\}$.
The optimality of $\hat \bsbB$ implies
\begin{align}
\frac{1}{2}\| \bsbX \hat \bsbB - \bsbX \bsbB^*\|_F^2 \leq \frac{1}{2}\| \bsbX \bsbB - \bsbX \bsbB^*\|_F^2 + A P_o(\bsbB) -A  P_o(\hat \bsbB)
+ \langle \bsbE, \bsbX \hat \bsbB - \bsbX \bsbB \rangle. \label{firstineq}
\end{align}
Let $\bsbDelta = \hat \bsbB - \bsbB$, $\hat{\mathcal J}= \mathcal J(\hat\bsbB)$, $\mathcal J= \mathcal J(\bsbB)$,   $J=J(\bsbB)$, $\hat J =  J(\hat\bsbB)$,  $r = r(\bsbB)$, $\hat r = r(\hat \bsbB)$.
Let  $\Proj_{rs}$  be  the orthogonal projection onto the  row space   of $\bsbX_{{\mathcal J}} \bsbB_{{\mathcal J}}$ and $\Proj_{rs}^{\perp}$ be its orthogonal complement.
Decompose $\bsbX \bsbDelta$ as follows
\begin{equation}\begin{aligned}
 \bsbX \bsbDelta = &\bsbX \bsbDelta \Proj_{rs} + \bsbX \bsbDelta \Proj_{rs}^{\perp} \\
= & \Proj_{\mathcal J} \bsbX \bsbDelta \Proj_{rs} + \Proj_{\mathcal J}^{\perp} \bsbX \bsbDelta \Proj_{rs} + \bsbX_{\hat{\mathcal J}} \hat\bsbB_{\hat{\mathcal J}} \Proj_{rs}^{\perp} \\
= & \Proj_{\mathcal J} \bsbX \bsbDelta \Proj_{rs} + \Proj_{\mathcal J}^{\perp} \bsbX_{\hat{\mathcal J}} \hat\bsbB_{\hat{\mathcal J}} \Proj_{rs} + \bsbX_{\hat{\mathcal J}} \hat\bsbB_{\hat{\mathcal J}} \Proj_{rs}^{\perp}.
\end{aligned}\label{decompofXD}
\end{equation}
Clearly,
$\| \bsbX \bsbDelta \|_F^2  = \| \Proj_{\mathcal J} \bsbX \bsbDelta \Proj_{rs}\|_F^2 + \|\Proj_{\mathcal J}^{\perp} \bsbX_{\hat{\mathcal J}} \hat\bsbB_{\hat{\mathcal J}} \Proj_{rs}\|_F^2 + \|\bsbX_{\hat{\mathcal J}} \hat\bsbB_{\hat{\mathcal J}}  \Proj_{rs}^{\perp}\|_F^2$.
It follows that
\begin{align}
\langle \bsbE, \bsbX \bsbDelta  \rangle &=  \langle \bsbE, \Proj_{\mathcal J} \bsbX \bsbDelta \Proj_{rs} \rangle + \langle \bsbE, \Proj_{\mathcal J}^{\perp} \bsbX_{\hat{\mathcal J}} \hat\bsbB_{\hat{\mathcal J}}\Proj_{rs}\rangle + \langle \bsbE, \bsbX_{\hat{\mathcal J}} \hat\bsbB_{\hat{\mathcal J}} \Proj_{rs}^{\perp}\rangle\notag\\&\equiv  \ \mbox{\textit{I}} + \mbox{\textit{II}} + \mbox{\textit{III}}. \label{decomp4}
\end{align}

\begin{lemma}\label{emprocBnd}
Suppose $\vect(\bsbE)$ is  sub-Gaussian with mean zero and  $\psi_2$-norm bounded by $\sigma$.  
Given $\bsbX\in \mathbb R^{n\times p}$, $1\leq J\leq p$, $1\leq r \leq J\wedge m$, define $\Gamma_{J, r} = \{\bsbDelta\in \mathbb R^{n\times m}: \|\bsbDelta\|_F\leq 1, r(\bsbDelta) \leq r, CS(\bsbDelta) \subset CS(\bsbX_{\mathcal J}) \mbox{ for some } \mathcal J: | \mathcal J|=J\}$. Let $$P_o'(J, r) = \sigma^2 \left\{(q\wedge  J ) r+(m - r) r + \log {p\choose J}\right\}.$$
Then for any $t\geq 0$,
\begin{align}
\EP [\sup_{\bsbDelta \in \Gamma_{J,r}} \langle \bsbE, \bsbDelta \rangle \geq t \sigma +  \{L  P_o'(J,r)\}^{1/2} ] \leq C\exp(- ct^2),
\end{align}
where $L, C, c>0$ are universal constants.
\end{lemma}

Term  \textit{I} and  term \textit{III}  on the right-hand side of \eqref{decomp4} can   be handled by Lemma \ref{emprocBnd}. Take the first term as an instance. For any $a, b, a'>0$,
\begin{align*}
&  \langle \bsbE, \Proj_{\mathcal J} \bsbX \bsbDelta \Proj_{rs} \rangle - \frac{1}{a} \|\Proj_{\mathcal J} \bsbX \bsbDelta \Proj_{rs}\|_F^2 -  b L P_o(J, r)   \\
\leq &
\|\Proj_{\mathcal J} \bsbX \bsbDelta \Proj_{rs}\|_F  \langle \bsbE, \Proj_{\mathcal J} \bsbX \bsbDelta \Proj_{rs} / \|\Proj_{\mathcal J} \bsbX \bsbDelta \Proj_{rs}\|_F\rangle - 2({{  b}/{a}})^{1/2} \|\Proj_{\mathcal J} \bsbX \bsbDelta \Proj_{rs}\|_F \{{L P_o(J, r)\}^{1/2}}  \\
\leq &
\frac{1}{a'} \|\Proj_{\mathcal J} \bsbX \bsbDelta \Proj_{rs}\|_F^2 +  \frac{a'}{4}
\sup_{1\leq J \leq p, 1\leq r \leq m\wedge J}  \sup_{\bsbDelta\in \Gamma_{J, r}} [\langle \bsbE, \bsbDelta \rangle - 2(b/a)^{1/2} \{{L P_o(J, r)\}^{1/2}}]_+^2\\
\equiv & \frac{1}{a'} \|\Proj_{\mathcal J} \bsbX \bsbDelta \Proj_{rs}\|_F^2 +  \frac{a'}{4} \sup_{1\leq J \leq p, 1\leq r \leq m\wedge J} R_{J,r}^2  \equiv  \frac{1}{a'} \|\Proj_{\mathcal J} \bsbX \bsbDelta \Proj_{rs}\|_F^2 +  \frac{a'}{4} R^2.
\end{align*}
Then,    with $4b >a$, we can   show that  $ R^2  $ is bounded by a constant in expectation. In fact,
\begin{align*}
& \EP(R\geq t \sigma ) \\
\leq & \sum_{J=1}^p \sum_{r=1}^{m \wedge J} \EP(R_{J,r}\geq t\sigma )\\
 \leq & \sum_{J=1}^p \sum_{r=1}^{m \wedge J}  \EP [\sup_{\bsbDelta \in \Gamma_{J,r}}  \langle \bsbE, \bsbDelta \rangle -\{L  P_o'(J,r)\}^{1/2} \geq t \sigma +   2(b/a)^{1/2} \{{L P_o(J, r)\}^{1/2}}-\{{L  P_o'(J,r)}\}^{1/2} ]\\
 \leq & \sum_{J=1}^p  \sum_{r=1}^{m \wedge J} C\exp(-c t^2) \exp[- c \{(2(b/a)^{1/2}-1)^2 L  P_o(J, r)/\sigma^2\}] \\
 \leq &  C  \exp(-c t^2) \exp(-c\log p)\sum_{J=1}^p\sum_{r=1}^{m \wedge J} \exp\{- c  L  P_o(J, r)/\sigma^2\}  \\
 \leq & C \exp(-c t^2) p^{-c},
\end{align*}
with the third inequality due to the Stirling formula,  the fourth inequality   due to the fact  that  $ J\log(ep/J) \geq \log p $ for any $J\geq 1$, and the fifth based on  the sum of geometric series. It follows from  $\EP(R\ge t\sigma) \le  C \exp(-c t^2)$ that $\EE( R^2) \le C \sigma^2$.

Similarly, for term \textit{III}, noticing that $r(\bsbX_{\hat{\mathcal J}} \hat\bsbB_{\hat{\mathcal J}} \Proj_{rs}^{\perp})\le r(\hat\bsbB_{\hat{\mathcal J}}) \le \hat r$ and $CS(\bsbX_{\hat{\mathcal J}} \hat\bsbB_{\hat{\mathcal J}} \Proj_{rs}^{\perp})\subset CS(\bsbX_{\hat {\mathcal J}})$, we have $$\langle \bsbE, \bsbX_{\hat{\mathcal J}} \hat\bsbB_{\hat{\mathcal J}} \Proj_{rs}^{\perp}\rangle-  \|\Proj_{\hat {\mathcal J}} \bsbX \bsbDelta \Proj_{rs}^{\perp}\|_F^2/a -  b L P_o(\hat J,\hat r) \le  \|\Proj_{\mathcal J} \bsbX \bsbDelta \Proj_{rs}\|_F^2/{a'} +  {a'} R^2/4.$$

\begin{lemma}\label{emprocBnd2}
Suppose $\vect(\bsbE)$ is  sub-Gaussian with mean zero and  $\psi_2$-norm bounded by $\sigma$.  
Given $\bsbX\in \mathbb R^{n\times p}$, $1\leq J\leq p$, $1\le J'\le p$, $1\leq r \leq J\wedge m$, define $\Gamma_{J',J, r} = \{\bsbDelta\in \mathbb R^{n\times m}: \|\bsbDelta\|_F\leq 1, r(\bsbDelta) \leq r,  CS(\bsbDelta) \subset CS(\Proj_{{\mathcal J}'}^{\perp} \Proj_{{\mathcal J}})   \mbox{ for some }  {\mathcal J}', {\mathcal J} \subset [p] \mbox{ satisfying }  | {\mathcal J}'|=J',  | {\mathcal J}|=J \}$. Let $$P_o''(J',J, r) = \sigma^2 \left[\{q\wedge  J \wedge (p- J')\} r  +(m - r) r + \log {p\choose J} + \log {p\choose J'}\right].$$
Then for any $t\geq 0$,
\begin{align*}
\EP [\sup_{\bsbDelta \in \Gamma_{J',J,r}} \langle \bsbE, \bsbDelta \rangle \geq t \sigma +   \{L  P_o''(J',J,r)\}^{1/2}] \leq C\exp(- ct^2),
\end{align*}
where $L, C, c>0$ are universal constants.
\end{lemma}

Likewise, we can get the following  bound for   \textit{II} from Lemma \ref{emprocBnd2}: 
\begin{align*}
 \langle \bsbE, \Proj_{\mathcal J}^{\perp} \bsbX_{\hat{\mathcal J}} \hat\bsbB_{\hat{\mathcal J}}\Proj_{rs}\rangle - \frac{1}{a} \|\Proj_{\mathcal J}^{\perp}\bsbX_{\hat{\mathcal J}} \hat\bsbB_{\hat{\mathcal J}}\Proj_{rs}\|_F^2-  b L P_o''( J, \hat J, \hat r)   \le
 \frac{1}{a'} \|\Proj_{\mathcal J}^{\perp} \bsbX \bsbDelta \Proj_{rs}\|_F^2 +  \frac{a'}{4} {R'}^2, \end{align*}
and       $\EE ({R'}^2) \leq C \sigma^2$ for some $L$ large enough.
In summary, we have
\begin{align*}
&\langle \bsbE, \bsbX \bsbDelta \rangle
\\ \leq & (\frac{1}{a}+\frac{1}{a'}) \|\bsbX \bsbDelta\|_F^2  + b L \{P_o(J,  r) + P_o(J, r) + P_o(\hat J, \hat r) +  P_o(\hat J,    \hat r)\}  + \frac{a'}{4} ( 2 R^2 +  {R'}^2)\\
\leq & (\frac{1}{a}+\frac{1}{a'})(1+b') \|\bsbX \bsbB - \bsbX \bsbB^*\|_F^2 +  (\frac{1}{a}+\frac{1}{a'})(1+\frac{1}{b'}) \|\bsbX \hat \bsbB - \bsbX \bsbB^*\|_F^2 \\ & + 2 b L \{P_o(J,   r) + P_o(\hat J,    \hat r)\}  +  \frac{a'}{4} ( 2 R^2 +  {R'}^2),
\end{align*}
where we used  $P_o''( J, \hat J, \hat r) \leq   P_o(J, r) + P_o(\hat J, \hat r)$.
Choosing constants   $a$, $a'$, $b$, $b'$, and $A$ sufficiently large such that $({1}/{a}+{1}/{a'})(1+{1}/{b'})<{1}/{2}$,  $4  b>a$,
and  $A > 2 b  L$,  we obtain the oracle inequality as desired.

{\remark Theorem 1 of  \cite{BSW12} gives a related but different result. In comparison, our penalty is of a smaller order and achieves the optimal rate. There is also  an error  in their   proof of Theorem 1:  
   $\|\bsbX_{\mathcal J \cap \hat{\mathcal J}} (\hat \bsbB - \bsbB)_{\mathcal J \cap \hat{\mathcal J}}\|_F^2 + \|\bsbX_{{\mathcal J}^c \cap \hat{\mathcal J}} (\hat \bsbB - \bsbB)_{{\mathcal J}^c \cap \hat{\mathcal J}}\|_F^2 + \|\bsbX_{\mathcal J \cap \hat{\mathcal J}^c} (\hat \bsbB - \bsbB)_{\mathcal J \cap \hat{\mathcal J}^c}\|_F^2 \lesssim \|\bsbX(\hat \bsbB - \bsbB)\|_F^2$ does {not} hold in general.
Our approach   is not subject to  the issue and results in a stronger conclusion.
}


%
%


\subsection{Proof of Theorem \ref{thm:sf-pic}}
\label{appsubsecSFPIC}

Let $h(\bsbB) = 1/\{mn - A P_o (\bsbB)/\sigma^2\}$.
By definition, $\hat \bsbB$ satisfies
\begin{align*}
& \| \bsbX \hat \bsbB - \bsbX \bsbB^*\|_F^2 \\
\leq & \|\bsbE\|_F^2 \{{h(\bsbB^*)}/{h(\hat\bsbB)}-1\} + 2\langle \bsbE, \bsbX \hat \bsbB - \bsbX \bsbB^*\rangle \\
 =&  \frac{A  \|\bsbE\|_F^2}{mn\sigma^2 - A P_o(\bsbB^*)} P_o(\bsbB^*) - \frac{A  \|\bsbE\|_F^2}{mn \sigma^2- A P_o(\bsbB^*)}P_o(\hat\bsbB) + 2\langle \bsbE, \bsbX \hat \bsbB - \bsbX \bsbB^*\rangle
\\
\leq & \frac{A  \|\bsbE\|_F^2}{mn\sigma^2 - A P_o(\bsbB^*)} P_o(\bsbB^*) - \frac{A  \|\bsbE\|_F^2}{mn \sigma^2}P_o(\hat\bsbB) + 2\langle \bsbE, \bsbX \hat \bsbB - \bsbX \bsbB^*\rangle.
\end{align*}

The stochastic term $ 2\langle \bsbE, \bsbX \hat \bsbB - \bsbX \bsbB^*\rangle$ can be decomposed and bounded in the same way as in the proof of Theorem \ref{thm:oracle}, except that     we use the  high-probability form  results here. For example, for   term $\langle \bsbE, \Proj_{\mathcal J} \bsbX \bsbDelta \Proj_{rs} \rangle$   in \eqref{decomp4},  Lemma \ref{emprocBnd} shows  that for any constants $a, b, a'>0$ satisfying $4b >a$, the following event
\begin{align*}
&  \langle \bsbE, \Proj_{\mathcal J} \bsbX \bsbDelta \Proj_{rs} \rangle \leq  ({1}/{a} +{1}/{a'}) \|\Proj_{\mathcal J} \bsbX \bsbDelta \Proj_{rs}\|_F^2 +  b L P_o(J, r)
\end{align*}
 occurs with   probability at least
$
1-  \sum_{J=1}^p \sum_{r=1}^{m \wedge J} C  \exp[- c \{(2({{b}/{a}})^{1/2}-1\}^2 L P_o(J, r)/\sigma^2]$ or $1-C p^{-c}$ for a sufficiently large value of $L$. Making use the fact and repeating the argument in the proof of   Theorem \ref{thm:oracle},   the following event
$$
 2\langle \bsbE, \bsbX  \hat \bsbB - \bsbX \bsbB^* \rangle
\leq2  ({1}/{a}+{1}/{a'})  \|\bsbX \hat \bsbB - \bsbX \bsbB^*\|_F^2   + 4b L \{P_o(\hat\bsbB) + P_o(\bsbB^*)\}
$$
occurs with probability at least $1-C p^{-c}$ for some $c, C>0$.

Let $\gamma$ and $\gamma'$ be  constants satisfying $0< \gamma <\ 1, \gamma'>0$. On $\mathcal A =\{(1-\gamma) {mn\sigma^2}\leq \|\bsbE\|_F^2 \leq (1+\gamma') {mn\sigma^2} \}$ ,  we have
$$
\frac{A  \|\bsbE\|_F^2}{mn\sigma^2 - A P_o(\bsbB^*)} P_o(\bsbB^*) - \frac{A  \|\bsbE\|_F^2}{mn \sigma^2}P_o(\hat\bsbB) \leq   \frac{(1+\gamma') A A_0}{A_0 - A}P_o(\bsbB^*)-   {(1-\gamma) A }P_o(\hat\bsbB).
$$
From \cite{lm2000}, the complement of $\mathcal A$ occurs with probability at most $C'\exp(-c' m n)$  with $c', C'$ dependent on $\gamma, \gamma'$.
 When $A_0$ is large enough, we can choose  $a, a', b, A$ such that    $({1}/{a}+{1}/{a'})<{1}/{2}$,  $4  b>a$, and $4b L\leq ( 1-\gamma)A$, and the conclusion results.

\subsection{Proof of Theorem \ref{thm:algconv}}
\label{app:subsec:algconv}
We first show the function-value convergence.
%
Given $\bsbV$, the $\bsbS$-optimization  reduces to a group variable selection problem
\begin{align*}
\frac{1}{2K} \|\vect( (\bsbY \bsbV)^{\tran}) - ( \bsbX \otimes  \bsbI) \vect(\bsbS^{\tran})\|_F^2 + \sum_{j=1}^p P(\|\bsbs_j\|_2;\lambda), 
\end{align*}
where  $\otimes$ is the Kronecker product.
Applying  Theorem 1 of \cite{SheGLMTISP} gives  the following result.

\begin{lemma} \label{senstS}
Let $\Theta$ be an arbitrary threshold function and $P$ be a $\Theta$-associated penalty as in the theorem.
Given  $\bsbV\in {\mathbb O}^{m\times r}$ and  $\bsbS\in {\mathbb R}^{p\times r}$, define
$T_{\bsbV}: {\mathbb R}^{p\times r} \rightarrow {\mathbb R}^{p\times r}$ by
$T_{\bsbV} \circ \bsbS =  \vec\Theta\left(\bsbXi;  \lambda  \right)$ with  $$\bsbXi=(\bsbxi_1, \ldots, \bsbxi_p)^{\tran}=\bsbX^{\tran}\bsbY\bsbV/K  + (\bsbI -  \bsbX^{\tran} \bsbX/K) \bsbS.$$ Assume $K\geq \|\bsbX\|_2^2$. Then,  $F(\bsbS, \bsbV) \ge F(T_{\bsbV} \circ  \bsbS, \bsbV)$, and if $\Theta$ is continuous at $\|\bsbxi_1\|_2, \allowbreak\ldots, \allowbreak \|\bsbxi_p\|_2$,
$F(\bsbS, \bsbV) - F(T_{\bsbV} \circ  \bsbS, \bsbV) \geq ({ 1-\|\bsbX\|_2^2/K}) \| T_{\bsbV} \circ  \bsbS- \bsbS \|_F^2/2$.
\end{lemma}

It follows from Lemma  \ref{senstS} and the triangle inequality that the inner loop of the algorithm  satisfies
\begin{align*}F(\bsbS^{(t-1)}, \bsbV^{(t)}) - F( \bsbS^{(t)}, \bsbV^{(t)}) & \geq { (1-\|\bsbX\|_2^2/K)} \sum_{l=1}^{\alpha_t}\|   {\tilde \bsbS}^{(l-1)} - {\tilde \bsbS}^{(l)} \|_F^2/2 \\ &\geq  { (1-\|\bsbX\|_2^2/K)} \|   \bsbS^{(t-1)} - \bsbS^{(t)} \|_F^2/2.\end{align*}

Given $\bsbS$, Steps (3)--(4) give a  solution to the $\bsbV$-minimization  over ${\mathbb O}^{m\times r}$    by  Procrustes rotation;  see, e.g.,   \cite{SheTISPMat} for a justification.
Hence for any $t\geq 1$, $F(\bsbS^{(t-1)}, \bsbV^{(t-1)})\geq  F(\bsbS^{(t-1)}, \bsbV^{(t)})\geq F(\bsbS^{(t)}, \bsbV^{(t)}) + { (1-\|\bsbX\|_2^2/K)} \| \bsbS^{(t-1)} - \bsbS^{(t)} \|_F^2/2$.

Next,  we prove the  accumulation-point property.   Observe that in updating $\bsbS$, the algorithm allows one to perform $T_{\bsbV^{(t)}}$ for $\alpha_t$  many   times, provided that $\alpha_t$ does not go beyond the pre-specified $M_{\mbox{\footnotesize inner}}$. To handle the ambiguity of $\bsbS^{(t)}$ and the non-uniqueness of $\bsbV^{(t)}$, we  use  {point-to-set} mappings \citep{Luenberger08}.
Let $T_{\bsbV}^\alpha$ with  $\alpha\ge 1$  be the composition of $\alpha$ $T_{\bsbV}$'s.
Define      ${\mathcal M}^{S}(\bsbS, \bsbV) = \{(\bar \bsbS, \bar \bsbV)\in \Omega: \bar \bsbS \in \{ T_{\bsbV} \circ \bsbS, T_{\bsbV}^2  \circ \bsbS, \ldots,  T_{\bsbV}^{M_{\mbox{\footnotesize inner}}} \circ \bsbS\}, \bar\bsbV=\bsbV\}$,
and  ${\mathcal M}^V(\bsbS, \bsbV) = \{(\bar\bsbS, \bar \bsbV)\in \Omega: \inf_{\tilde \bsbV \in \mathbb{O}^{m\times r}} F(\bsbS, \tilde \bsbV) \geq F(\bsbS, \bar \bsbV), \bar \bsbS=\bsbS\}$.
Then     $\mathcal{M}= {\mathcal M}^S{\mathcal M}^V $  characterizes  the algorithm.

By the continuity assumption, it is easy to show that $\mathcal M$ is {closed} at any accumulation point of $\{\bsbS^{(t-1)}, \bsbV^{(t)}\}$, 
 cf. Lemma 14 in \citealt{BSW12}. Moreover, from
 $F(\bsbS^{(0)}, \bsbV^{(1)})-F(\bsbS^{(t)}, \bsbV^{(t)})\geq { (1-\|\bsbX\|_2^2/K)} \| \bsbS^{(0)} - \bsbS^{(t)} \|_F^2/2$,
$\bsbS^{(t)}$, $\bsbV^{(t)}$, and  $\bsbB^{(t)}$ are all  uniformly bounded in $t$.

The rest of the proof proceeds along  similar lines of the proof of Theorem 7 in \cite{BSW12}, with minor modifications.  Let $(\bsbS^*, \bsbV^*)\in \Omega$ be an arbitrary accumulation point   of $(\bsbS^{(t-1)}, \bsbV^{(t)})$. Assume, without loss of generality, that  $(\bsbS^{(t)}, \bsbV^{(t+1)})$ converges to  $(\bar \bsbS, \bar \bsbV)$. Then $(\bar \bsbS, \bar \bsbV)\in {\mathcal M}(\bsbS^*, \bsbV^*)$,  since $\mathcal M$ is closed. It follows from the first conclusion that     $F(\bar \bsbS, \bar \bsbV) = F(\bsbS^*, \bsbV^*)$, and thus $F(\bar \bsbS, \bsbV^*) = F(\bsbS^*, \bsbV^*)$.   Lemma \ref{senstS} further implies  $\bar \bsbS = \bsbS^*$, or $T_{\bsbV^*}^{\alpha_0} \circ \bsbS^* = \bsbS^*$ for some $\alpha_0\in {\mathbb N}$. So $F(T_{\bsbV^*}^{\alpha} \circ \bsbS^*, \bsbV^*) = F(\bsbS^*, \bsbV^*)$ for any $\alpha \leq \alpha_0$. In particular, $F(T_{\bsbV^*} \circ \bsbS^*, \bsbV^*) = F(\bsbS^*, \bsbV^*)$.
Applying Lemma \ref{senstS} again yields
$T_{\bsbV^*} \circ \bsbS^* = \bsbS^*$, i.e.,   $\bsbS^*$ is a fixed point of $T_{\bsbV^*}$. Following the lines of  the proof of Lemma 1 in \cite{SheGLMTISP},  it can be shown that for fixed $(\bsbs_1, \ldots, \bsbs_{j-1}, \bsbs_{j+1},\ldots, \bsbs_p$, $\bsbV^*)$, $\bsbs_j^*$ is a global minimizer of  $F(\bsbs_1, \ldots, \bsbs_j, \ldots, \bsbs_p, \bsbV^*)$, $j=1,\ldots, p$.
On the other hand, from $\bar \bsbS = \bsbS^*$, we have $(\bsbS^*, \bar \bsbV)  \in {\mathcal M}^V(\bsbS^*, \bsbV^*)$ and   $\bar \bsbV$ is a  global  minimizer of $F(\bsbS^*, \bsbV)$ given $\bsbS^*$. But $F(\bsbS^*, \bar \bsbV) = F(\bar \bsbS, \bar \bsbV) = F(\bsbS^*, \bsbV^*)$, so $\bsbV^*$ also  minimizes $F(\bsbS^*, \bsbV)$ given $\bsbS^*$.
The proof is complete.

{\remark A more careful study following  \cite{SheGLMTISP}  can be made to strengthen  the factor $(1-\|\bsbX\|_2^2/K)$  to
$1+\essinf$ $\{ \rd \Theta^{-1}(u;\lambda)/\rd u:  u\ge 0\}-\|\bsbX\|_2^2/K$,
where $\Theta^{-1}(u;\lambda)=\sup\{s:\Theta(s;\lambda)\leq u\}$ and   $\essinf$ is the   essential infimum.
  For example,   when  $\Theta(s;\lambda)=\mbox{sgn}(s)(|s|-\lambda)_+$, which corresponds to the  $\ell_1$-penalty $P(\theta;\lambda)=\lambda |\theta|$,        $K=\|\bsbX\|_2^2/2$  suffices. 
}
%

\subsection{Proof of Theorem \ref{thm:algconv-scr}}
\label{app:subsec:algconv-scr}
%
Given any $\bsbXi \in \mathbb R^{p\times m}$,
$\hat \bsbB  = \vec\Theta^{\#}(\bsbXi; {d} , \eta)$ is an optimal solution to
\begin{align}
\min_{\bsbB\in \mathbb R^{p\times m}} \frac{1}{2} \| \bsbXi - \bsbB\|_2^2  + \frac{\eta}{2} \|\bsbB\|_2^2= f(\bsbB;\eta) \ \mbox{ subject to } J(\bsbB) \leq {d}. \label{protoopt1}
\end{align}
In fact,
for any given $\mathcal J\subset \mathcal [p]$ with $|\mathcal J|={d}$, if   $\bsbB({\mathcal J^c},1:m)=\bsb{0}$,  the optimal solution $\hat\bsbB$ must satisfy      $\hat\bsbB({\mathcal J},1:m)=\bsbXi(\mathcal J,1:m)/(1+\eta)$ and  $f(\hat \bsbB; \eta)=\|\bsbXi\|_F^2/2 - \|\bsbXi(\mathcal J, 1:m)\|_F^2/\{2(1+\eta)\} $. Hence  $\vec\Theta^{\#}(\bsbXi; {d}, \eta)$ delivers a global minimizer to \eqref{protoopt1}.

For  the   general problem which involves  a   predictor matrix,  the   $\bsbS$-optimization with $\bsbV$ held fixed amounts to  minimizing
$g(\bsbS)=\|\bsbY' - \bsbX \bsbS \|_F^2/(2K) + {\eta}\| \bsb{S} \|_F^2/2$ subject to  $\|\bsbS\|_{2,0}\leq {d}$, where   $\bsbY' = \bsbY \bsbV$. Similar to    Lemma \ref{senstS}, we can construct a   surrogate function   $G(\bsbS, \tilde\bsbS)=g(\tilde\bsbS) + \langle (\bsbI - \bsbX^{\tran} \bsbX/K) (\bsbS - \tilde \bsbS), \bsbS - \tilde \bsbS\rangle$, and apply an iterative quantile thresholding algorithm that  gurantees the function value is decreasing.   The rest of the proof can be shown following  the lines of the proof of Theorem \ref{thm:algconv};  the details are omitted.

\section{Proofs of Lemmas}


\subsection{Proof of Lemma \ref{lemma:phostochastic}}
\begin{proof}
Let   $l_H(\bsbDelta, r) =  2\langle \bsbE, \bsbX  \bsbDelta\rangle -  \|\bsbX  \bsbDelta \|_F^2/a - 2aA_0\{ P_{2,H}(\rho\bsbDelta; {\lambda^o}) + \sigma^2 r(m-r)\}$ and $l_0(\bsbDelta, r) =  2\langle \bsbE, \bsbX  \bsbDelta\rangle -  \|\bsbX  \bsbDelta \|_F^2/a - 2aA_0\{  P_{2,0}(\rho\bsbDelta; {\lambda^o}) + \sigma^2 r(m-r)\}$. Event    $\mathcal A$      as defined  by \eqref{event0}   can be formulated as   $$\sup_{r:1\le r \le m}  \sup_{\bsbDelta: r(\bsbDelta)\leq r} l_H(\bsbDelta, r)\geq at\sigma^2.$$
  Let     $\mathcal A_H=\{\sup_{\bsbDelta: r(\bsbDelta)\leq r} l_H(\bsbDelta, r)\geq at\sigma^2\}$ and $\mathcal A_0=\{\sup_{\bsbDelta: r(\bsbDelta)\leq r} l_0(\bsbDelta, r)\geq at\sigma^2\}$.
The occurrence of  $\mathcal A_H$ implies that 
\begin{align*}l_H(\bsbDelta^o, r) \geq a t \sigma^2 \ \mbox{ for any } \bsbDelta^o \in \arg \min_{r(\bsbDelta)\leq r}  \frac{1}{ a } \|\bsbX  \bsbDelta \|_F^2 -2\langle \bsbE, \bsbX  \bsbDelta\rangle +2a A_0  P_{2,H}(\rho\bsbDelta; {\lambda^o}).\end{align*}
From Lemma \ref{lemma:phcomp} and $a^2A_0\geq 1$,  there exists   $\bsbDelta^o$ such that $l_0(\bsbDelta^o, r)=l_H(\bsbDelta^o, r)  \geq a t \sigma^2$. Hence  $\mathcal A_H \subset \mathcal \mathcal A_0$ and it suffices to show $\EP (\mathcal A_0) \le C \exp(-ct)$. We point out that the optimization-based   reduction  is by no means trivial because $P_0(s; \lambda) \geq P_H(s; \lambda)$, for any $s\in \mathbb R$.

Applying von-Neumann's trace inequality and Cauchy-Schwarz inequality gives
\begin{align*}
&2\langle \bsbE, \bsbX \bsbDelta\rangle \\
=& 2\langle \Proj_{\bsbX_{{\mathcal J}(\bsbDelta)}} \bsbE \Proj_{RS(\bsbDelta)}, \bsbX \bsbDelta\rangle \\
\leq& 2\|\Proj_{\bsbX_{{\mathcal J}(\bsbDelta)}} \bsbE \Proj_{RS(\bsbDelta)}\|_F  \| \bsbX \bsbDelta \|_F \\
\leq& a \|\Proj_{\bsbX_{{\mathcal J}(\bsbDelta)}} \bsbE \Proj_{RS(\bsbDelta)}\|_F^2 + \frac{1}{a} \| \bsbX \bsbDelta \|_F^2.
\end{align*}
This means that $l_0(\bsbDelta, r)\ge  at \sigma^2 $ implies $ \|\Proj_{\bsbX_{{\mathcal J}(\bsbDelta)}} \bsbE \Proj_{RS(\bsbDelta)}\|_F^2-2A_0\{\sigma^2(r + \log p) J(\bsbDelta)  +\sigma^2 r(m-r)\} \ge t\sigma^2$.
Therefore,    $\EP(\mathcal A_0)$ is bounded above by
\begin{align*}
&\EP\{\sup_{r:1\le r \le m}  \sup_{\bsbDelta: r(\bsbDelta)\leq r} l_H(\bsbDelta, r)\geq at\sigma^2\} \\  = &   \EP\{\sup_{J:1\le J \le p}  \sup_{r:1\le r \le m\wedge J}\sup_{\bsbDelta: |\mathcal J(\bsbDelta)| = J, r(\bsbDelta)\leq r} l_H(\bsbDelta, r)\geq at\sigma^2\}\\
\le & \sum_{J=1}^p \sum_{r=1}^{m\wedge J} \EP[R_{J, r} \ge   2A_0\{\sigma^2(r + \log p) J  +\sigma^2 r(m-r)\} + t\sigma^2],
 \end{align*}
where we use  the notation $R_{J,r}$  in  Lemma \ref{concenGauss}.  Setting $A_0$ sufficiently large, e.g., $A_0\ge  L$ with $L$ the universal constant in  Lemma \ref{concenGauss}, leads to     $A_0\{\sigma^2(r + \log p) J  +\sigma^2 r(m-r)\} \ge L P_o(J,r) + c A_0 J r \sigma^2 + c A_0 \sigma^2\log p$ for some $c>0$. From  \eqref{chisqresbound} and the sum of geometric series, we obtain
\begin{align}
\EP(\mathcal E_0) &\le \sum_{J=1}^p \sum_{r=1}^{m\wedge J} \EP \{R_{J, r}\ge  L P_o(J, r) + c A_0 J r \sigma^2 +c A_0 \sigma^2\log p+ t\sigma^2\} \notag\\
 & \le \sum_{J=1}^p \sum_{r=1}^{m\wedge J} C \exp(-ct) \exp(-c A_0 Jr ) \exp(-c A_0 \log p)\notag\\
 &\le C \exp(-ct) p^{-c A_0}, \notag
\end{align}
and hence $\EP(\mathcal E_0) \le C \exp(-ct )$.  
\end{proof}

\subsection{Proof of Lemma \ref{lemma:phcomp}}
\begin{proof}
The proof  follows similar lines as the  proof of Theorem  \ref{thm:algconv} but does not   require the  continuity condition. Write $\bsbB = \bsbS \bsbV^{\tran}$ and define $F(\bsbB)$ or $F(\bsbS, \bsbV)$ as $\| \bsbY - \bsbX \bsbS \bsbV^{\tran}\|_F^2/2 +  P_{2,H}(\bsbS; \lambda)$.
Let $\bsbB_o = \bsbS_o  \bsbV_o^{\tran}$ be an arbitrary global minimizer of $F$. Construct $\bsbS^o= \vec\Theta_H\{\bsbX^{\tran}  \bsbY \bsbV_o + (\bsbI -  \bsbX^{\tran} \bsbX)  \bsbS_o; {\lambda}\}$,  where $\vec\Theta_H$ is the multivariate hard thresholding induced by  $\Theta_H(t; \lambda) = t 1_{|t|>\lambda}$. As in the proof of Theorem \ref{thm:algconv}, with a surrogate function    $G_{\bsbV}(\bsbS, \tilde\bsbS)=\|\bsbY \bsbV  - \bsbX \tilde\bsbS \|_F^2/2 +P_{2,H}(\tilde \bsbS; {\lambda}) +\langle (\bsbI - \bsbX^{\tran} \bsbX) (\bsbS - \tilde \bsbS), \bsbS - \tilde \bsbS\rangle/2 + \|\bsbY \bsbV_{\perp}\|_F^2/2$,  we can obtain  $$F(\bsbS_o, \bsbV_o) \ge G_{\bsbV_o}(\bsbS_o, \bsbS^o)\geq F(\bsbS^o, \bsbV_o),$$
with the first inequality due to the construction of $\bsbS^o$,   and the second inequality   due to $\|\bsbX\|_2\le 1$.   Therefore,  $\bsbB^o =\bsbS^o\bsbV_o^{\tran}$ is  a global minimizer of $F$ as  well. Clearly,  $\|\bsb{b}_{j}^o\|_2 =\| \bsbs_{j}^o\|_2$. From the definition of $\vec\Theta_H$, $\bsb{b}_j^o$ satisfy the desired property.
\end{proof}

\subsection{Proof of Lemma \ref{concenGauss}}
\begin{proof}
First,  for fixed  $\mathcal J$ and $\bsbV$, $R_{\mathcal J, \bsbV}/\sigma^2\sim \chi^2(df)$ with $df\leq (J\wedge q)r$. The  standard tail bound for $X\sim \chi^2(df)$ \cite[Lemma 1]{lm2000} gives  $\EP(X - df \geq t)\leq \exp[-{t^2}/\{4{(df + t)}\}]$  for any $ t\geq 0$  .

Let  $\mathcal T$ be  an $\varepsilon$-net of the  Grassmann  manifold  $\{\Proj_{\bsbV}: \bsbV \in \mathbb O^{m\times r}\}$ under the operator norm $\|\cdot\|_2$. From \cite{szarek82},  the cardinality of the $\varepsilon$-net ${\mathcal N}(\varepsilon)$ satisfies
$
{\mathcal N}(\varepsilon) \leq   \left({C}/{\varepsilon}\right)^{r(m-r)},
$
where   $C$ is a universal constant. Given  $\bsbV_1\in \mathbb O^{m\times r}$, there exists   $\bsbV_2$ with $\Proj_{\bsbV_2}\in \mathcal T$ such that $\| \bsbV_1 \bsbV_1^{\tran} - \bsbV_2 \bsbV_2^{\tran} \|_2\leq \varepsilon$. Let $\tilde \bsbU \tilde \bsbD \tilde V$ be the \ full  singular value decomposition of  $[\bsbV_1, \bsbV_2]$, where   $\tilde \bsbU=[\tilde \bsbU_1, \tilde \bsbU_2]$ with both   $\tilde\bsbU_1$  and $\tilde \bsbU_2$  having $r$ columns. Then
\begin{align*}
\|\Proj_{\bsbX_{\mathcal J}} \bsbE\bsbV_1\|_F^2 &= \mathrm{tr}(\Proj_{\bsbX_{\mathcal J}} \bsbE\bsbV_2 \bsbV_2^{\tran} \bsbE^{\tran}) + \mathrm{tr}\{\Proj_{ \bsbX_{\mathcal J}} \bsbE(\bsbV_1 \bsbV_1^{\tran}-\bsbV_2 \bsbV_2^{\tran}) \bsbE^{\tran}\}\\
 &= \mathrm{tr}(\Proj_{\bsbX_{\mathcal J}} \bsbE\bsbV_2 \bsbV_2^{\tran} \bsbE^{\tran}) +\mathrm{tr}\{\Proj_{\bsbX_{\mathcal J}} \bsbE \Proj_{\tilde\bsbU} (\bsbV_1 \bsbV_1^{\tran}-\bsbV_2 \bsbV_2^{\tran}) \Proj_{\tilde\bsbU} \bsbE^{\tran}\}\\
&\leq \mathrm{tr}(\Proj_{\bsbX_{\mathcal J}} \bsbE\bsbV_2 \bsbV_2^{\tran} \bsbE^{\tran}) +\varepsilon \| \Proj_{\bsbX_{\mathcal J}} \bsbE \Proj_{\tilde\bsbU}\|_F^2 \\
&=\mathrm{tr}(\Proj_{\bsbX_{\mathcal J}} \bsbE\bsbV_2 \bsbV_2^{\tran} \bsbE^{\tran}) +\varepsilon \| \Proj_{\bsbX_{\mathcal J}} \bsbE \Proj_{\tilde\bsbU_1}\|_F^2 + \varepsilon \| \Proj_{\bsbX_{\mathcal J}} \bsbE \Proj_{\tilde\bsbU_2}\|_F^2.
\end{align*}
Taking the supreme on both sides, we get
$
R_{J,r} \leq \max_{\mathcal J: |\mathcal J| = J, \Proj_{\bsbV}\in \mathcal T} \|\Proj_{\bsbX_{\mathcal J}} \bsbE\Proj_{\bsbV}\|_F^2 + 2 \varepsilon R_{J, r}$ or
$$
R_{J,r} \leq {\max_{\mathcal J: |\mathcal J| = J, \Proj_{\bsbV}\in \mathcal T} \|\Proj_{\bsbX_{\mathcal J}} \bsbE\Proj_{\bsbV}\|_F^2}/({1-2\varepsilon}).
$$
 Applying the $\chi^2$ tail bound and the union bound  gives
\begin{align*}
\EP  \left\{R_{J, r}  \geq \sigma^2\frac{(J\wedge q)r + t}{1-2\varepsilon}  \right\} \leq {p \choose J} \left(\frac{C}{\varepsilon}\right)^{r(m-r)} \exp[-Ct^2/\{t+(J\wedge q)r\}].
\end{align*}
The  conclusion follows  for  $L$ large enough.
\end{proof}

\subsection{Proofs of Lemma \ref{emprocBnd} and Lemma \ref{emprocBnd2}}
\begin{proof}
By definition, for any fixed $\bsbDelta$, $\langle \bsbE,  \bsbDelta\rangle$ is a mean-centered sub-Gaussian random variable with scale bounded by $\sigma\|\bsbDelta\|_F$. Therefore,   $\{\langle \bsbE,  \bsbDelta\rangle: \bsbDelta \in \Gamma_{J,r}\}$ is a stochastic process with sub-Gaussian increments.   Dudley's entropy integral can be used to bound its  supremum, see, e.g., \cite{talagrand2014upper}. The induced metric on $\Gamma_{J,r}$ is Euclidean: $d(\bsbDelta_1, \bsbDelta_2) = \sigma \|\bsbDelta_1 - \bsbDelta_2\|_F$.

To bound   the metric entropy $\log {\mathcal N}(\varepsilon, \Gamma_{J,r}, d)$, where ${\mathcal N}(\varepsilon, \Gamma_{J,r}, d)$ is the smallest cardinality of an $\varepsilon$-net that covers $\Gamma_{J,r}$  under  $d$, we  characterize   each matrix in $\Gamma_{J, r}$ using its row/column spaces, motivated by \cite{Recht2010}. Given $\bsbDelta\in \Gamma_{J,r}$,   its column space must be contained in  $CS(\bsbX_{\mathcal J})$ for some $\mathcal J$  with $|\mathcal J|=J$, and  its row space must be contained in an $r$-dimensional subspace in $\mathbb R^m$. Hence
\begin{align}
\bsbDelta  =\bsbU \bsbSig \bsbV^{\tran},\label{rcdecomp}
\end{align}
 where  $\Proj_{\bsbU}=\Proj_{\bsbX_{\mathcal J}}$, $\bsbV\in \mathbb O^{m\times r}$, $\bsbSig \in \mathbb R^{(J\wedge q)\times r}$.

It is easy to see that the number of candidate $\bsbX_{\mathcal J}$ is no more than the number of ways of choosing $J$ dimensions from $p$ dimensions, and     $RS(\bsbDelta)$ is a point on the Grassmann manifold  $G_{m, r}$ of all $r$-dimensional subspaces of $\mathbb R^m$. Equipped with   metric $d''$ that is     the operator norm, i.e.,  $\|\bsbV_1 \bsbV_1^{\tran} - \bsbV_2 \bsbV_2^{\tran} \|_2$ for any $\bsbV_1, \bsbV_2 \in \mathbb O^{m\times r}$,
$
{\mathcal N}(\varepsilon, G_{m,r}, d'') \leq   \left({C_1}/{\varepsilon}\right)^{r(m-r)}
$,
where   $C_1$ is a universal constant  \citep{szarek82}. Moreover, it is easy to see that $\bsbSig=\bsbU^{\tran} \bsbDelta \bsbV$ is in a unit ball of dimensionality ${(J\wedge q)\times r}$, denoted by $B_{(J\wedge q)\times r}$. By a standard volume argument, $${\mathcal N}(\varepsilon, B_{(J\wedge q)\times r}, d') \leq (C_0/\varepsilon)^{(J\wedge q)\times r},$$ where $d'$ is the Euclidean distance in $\mathbb R^{(J\wedge q)\times r}$ and   $C_0$ is a universal constant.  We claim that
under the metric $d$, \begin{align}
\log  {\mathcal N}(\varepsilon, \Gamma_{J,r}, d)  \leq  \log {p \choose J} + \{(J\wedge q) r +r(m-r)\}\log ({C \sigma}/{\varepsilon}).
\end{align}
In fact, given any $\bsbDelta_1\in \Gamma_{J,r}$, we can write  $\bsbDelta_1 = \bsbU_1 \bsbSig_1 \bsbV_1^{\tran}$ according to \eqref{rcdecomp} and  find $\bsbV_2$ and $\bsbSig_2$ such that $\|\bsbV_1\bsbV_1^{\tran} - \bsbV_2\bsbV_2^{\tran}\|_2 \leq \varepsilon$ and   $\|\bsbSig_1 \bsbV_1^{\tran} \bsbV_2 - \bsbSig_2\|_F\leq \varepsilon$.  Then,  for $\bsbDelta_2=\bsbU_1 \bsbSig_2 \bsbV_2^{\tran}$,
\begin{align*}
\|\bsbDelta_1 -\bsbDelta_2\|_F &\leq \|\bsbDelta_1 - \bsbDelta_1 \bsbV_2 \bsbV_2^{\tran}\|_F + \| \bsbU_1 \bsbSig_1 \bsbV_1^{\tran} \bsbV_2 \bsbV_2^{\tran} - \bsbU_1 \bsbSig_2 \bsbV_2^{\tran}\|_F \\&\leq [\mathrm{tr}\{ \bsbDelta_1^{\tran} \bsbDelta_1 (\Proj_{\bsbV_1}- \Proj_{\bsbV_2})^2\}]^{1/2} + \|  \bsbSig_1 \bsbV_1^{\tran} \bsbV_2 - \bsbSig_2 \|_F\\
&\leq  (\|\bsbDelta_1\|_F^2 \|\Proj_{\bsbV_1}- \Proj_{\bsbV_2}\|_2^2)^{1/2}+\varepsilon \leq 2\varepsilon.
\end{align*}

From Dudley's integral bound, we obtain
\begin{align*}
\EP \left\{\sup_{\bsbDelta \in \Gamma_{J,r}} \langle \bsbE, \bsbDelta \rangle \geq t \sigma + L \int_0^{\sigma}  {\log^{1/2}  {\mathcal N}(\varepsilon, \Gamma_{J,r}, d)}\rd \varepsilon\right\} \leq C\exp(- ct^2).
\end{align*}
Simple computation   gives
\begin{align*}
\int_0^{\sigma}  \log^{1/2} {\mathcal N}(\varepsilon, \Gamma_{J,r}, d) \rd \varepsilon &\lesssim \sigma {\log^{1/2}{p\choose J}} + \sigma \{(J\wedge q)\times r +r(m-r)\}^{1/2}\lesssim {P_o'^{1/2}(J, r)}.
\end{align*}
 The proof of  Lemma \ref{emprocBnd2} is similarly  based on the entropy integral bound. The details are omitted.
 \end{proof}

\bibliographystyle{apalike}
\bibliography{selectivefactor}

\end{document}